\documentclass[twocolumn]{aastex62}
\usepackage{bm}
\usepackage{amsmath}
\usepackage{booktabs}
\usepackage{graphicx}
\usepackage{comment}

\newcommand{\given}{\,|\,}
\newcommand{\dd}{\mathrm{d}}

\newcommand{\Teff}{\ifmmode {T_{\rm eff} }\else$T_{\rm eff}$\fi}

\newcommand{\degree}{\ifmmode {^{\circ}\ }\else$^{\circ}$\fi}
\newcommand{\amin}{\ifmmode {^{\prime}\ }\else$^{\prime}$\fi}
\newcommand{\asec}{\ifmmode {^{\prime\prime}}\else$^{\prime\prime}$\fi}

\newcommand{\ergs}{\ifmmode { {\rm erg}\ {\rm s}^{-1} } \else erg s$^{-1}$\fi}

\newcommand{\Msun}{\ifmmode {M_{\odot}}\else${M_{\odot}}$\fi}
\newcommand{\Rsun}{\ifmmode {R_{\odot}}\else${M_{\odot}}$\fi}
\newcommand{\Porb}{\ifmmode {P_{\rm orb}}\else${P_{\rm orb}}$\fi}

\newcommand{\gaia}{{\it Gaia}}

\submitjournal{AAS Journals}

\shorttitle{Wide Binaries in APOGEE}
\shortauthors{Andrews et al.}

\begin{document}

\title{Using APOGEE Wide Binaries to Test Chemical Tagging with Dwarf Stars}

\correspondingauthor{Jeff J.\ Andrews}
\email{andrews@physics.uoc.gr}

\author[0000-0001-5261-3923]{Jeff J.\ Andrews}
\affiliation{Foundation for Research and Technology-Hellas, 
100 Nikolaou Plastira St., 
71110 Heraklion, Crete, Greece}
\affiliation{Physics Department \& Institute of Theoretical \& Computational Physics, 
P.O Box 2208, 
71003 Heraklion, Crete, Greece}
\affiliation{Niels Bohr Institute, 
Blegdamsvej 17, 
2100 K\o benhavn \O, Denmark}

\author[0000-0001-5261-4336]{Borja Anguiano}
\affiliation{Department of Astronomy, 
University of Virginia, 
Charlottesville, VA, 22904, USA}
\affiliation{Department of Physics $\&$ Astronomy, 
Macquarie University, Sydney, 22904, Australia}

\author[0000-0003-2481-4546]{Julio Chanam\'e}
\affiliation{Instituto de Astrof\'isica, 
Pontificia Universidad Cat\'olica de Chile, 
Av.~Vicu\~na Mackenna 4860, 
782-0436 Macul, Santiago, Chile}
\affiliation{Millennium Institute of Astrophysics, 
Santiago, Chile}

\author[0000-0001-7077-3664]{Marcel A.\ Ag\"ueros}
\affiliation{Department of Astronomy, 
Columbia University, 
550 West 120th Street, 
New York, NY 10027, USA}

\author[0000-0002-7871-085X]{Hannah M.\ Lewis}
\affiliation{Department of Astronomy, 
University of Virginia, 
Charlottesville, VA, 22904, USA}

\author[0000-0003-2969-2445]{Christian R.\ Hayes}
\affiliation{Department of Astronomy, 
University of Virginia, 
Charlottesville, VA, 22904, USA}

\author[0000-0003-2025-3147]{Steven R.\ Majewski}
\affiliation{Department of Astronomy, 
University of Virginia, 
Charlottesville, VA, 22904, USA}

\begin{abstract}
Stars of a common origin are thought to have similar, if not nearly identical, chemistry. Chemical tagging seeks to exploit this fact to identify Milky Way subpopulations through their unique chemical fingerprints. In this work, we compare the chemical abundances of dwarf stars in wide binaries to test the abundance consistency of stars of a common origin. Our sample of 31 wide binaries is identified from a catalog produced by cross-matching APOGEE stars with UCAC5 astrometry, and we confirm the fidelity of this sample with precision parallaxes from \gaia\ DR2. For as many as 14 separate elements, we compare the abundances between components of our wide binaries, finding they have very similar chemistry (typically within 0.1 dex). This level of consistency is more similar than can be expected from stars with different origins (which show typical abundance differences of 0.3-0.4 dex within our sample). For the best measured elements, Fe, Si, K, Ca, Mn, and Ni, these differences are reduced to 0.05-0.08 dex when selecting pairs of dwarf stars with similar temperatures. Our results suggest that APOGEE dwarf stars may currently be used for chemical tagging at the level of $\sim$0.1 dex or at the level of $\sim$0.05 dex when restricting for the best-measured elements in stars of similar temperatures. Larger wide binary catalogs may provide calibration sets, in complement to open cluster samples, for on-going spectroscopic surveys.
\end{abstract}


\section{Introduction}
\label{sec:intro}

Wide binaries are bound pairs of stars separated by as much as $\sim$1 parsec. Several formation scenarios exist for these systems, including the dynamical unfolding of triple systems \citep{reipurth12}, association of stellar pairs during the dissolution of stellar clusters \citep{kouwenhoven10,moeckel11}, turbulent fragmentation \citep{lee17}, and the gravitational attraction of nearby pre-stellar cores \citep{tokovinin17}. All of these scenarios imply that wide binaries should be born at roughly the same time, from very similar pre-stellar material. The co-eval and co-chemical nature of these pairs allows them to be used for a variety of unique astrophysical applications, such as the calibration of otherwise difficult-to-assess M-dwarf metallicities \citep{lepine07,rojas_ayala10,montes18}, of the age-magnetic activity relation of stars \citep{garces11,chaname12}, of the age-metallicity relation \citep{rebassa-mansergas16}, of the initial-final mass relation for white dwarfs \citep{zhao12, andrews15}, and for tests of the presence and nature of dark matter in the Milky Way \citep{bahcall85, yoo04} and in ultra-faint dwarf galaxies \citep{penarrubia16}.

The co-eval and co-chemical nature of wide binaries has been largely justified on theoretical grounds; only a handful of studies have been performed demonstrating this premise observationally. \citet{makarov08} provided evidence that wide binaries are co-eval, and \citet{kraus09} showed that wide binaries in the Tauris-Auriga association have ages more consistent than the association as a whole. More work has focused on the chemical composition of wide binaries. Initial studies by \citet{gizis97} and \citet{gratton01} demonstrated the overall consistency in the metallicity of a handful of wide binaries. \citet{desidera04, desidera06b} focused on iron and found that the components of a sample 50 wide binaries typically show [Fe/H] measurements consistent to within 0.02 dex. Rarely did these studies extend beyond Fe abundances; exceptions include \citet{desidera06b} who additionally study vanadium and \citet{martin02} who focus on lithium. 

Separately, the elemental abundances of a number of wide binaries that host planets have been studied in detail using high-resolution spectra. These include X0-2N/X0-2S \citep{teske13,teske15}, 16 Cyg A/B \citep{laws01,ramirez11,schuler11a,tucci14}, HAT-P-1 \citep{liu14}, HD 20872/20871 \citep{mack14}, and HD 80606/80607 \citep{mack16, liu18}. Additionally, as part of a large spectroscopic study, \citet{brewer16} discuss the detailed abundance differences of nine wide binaries. These studies often find small but significant differences in the detailed abundance patterns of wide-binary components hosting planets, which has been attributed to the accretion of planetesimals \citep{mack14}. However, accretion is only one source of the elemental abundance differences between stars with initially identical chemistry; dredge-up, mixing, and radiative levitation can all act to alter the surface elemental abundances of a star \citep{schuler11a, dotter17,souto18}. Furthermore, uncertainties in fundamental stellar parameters such as \Teff\ and log $g$ can alter the apparent abundances of stars \citep[e.g.,][]{teske15}. Detailed study of putatively co-chemical stars, such as wide binaries, that span a significant age range can help determine the extent to which these various effects alter the chemistry of stars throughout their lifetimes.

In \citet{andrews18}, hereafter Paper I, we cross-matched our sample of wide binaries from \citet{andrews17} that were identified in the Tycho-\gaia\ Astrometric Solution \citep[TGAS;][]{lindegren16} with data from the Large Sky Area Multi-Object Fibre Spectroscopic Telescope \citep[LAMOST;][]{luo15} and the Radial Velocity Experiment \citep[RAVE;][]{kunder17} spectroscopic surveys. Using the combined sample of 177 wide binaries, in Paper I we argued that wide binaries have Fe abundances consistent within measurement uncertainties, which supports the premise that wide binary-components have a common origin. 

In Paper I we further suggested that wide binaries can be used to test chemical tagging, the process of identifying co-eval stellar subpopulations in the Galaxy using chemistry. This is based on the idea that stars of a common origin share unique chemical fingerprints \citep{freeman02}. Initial tests of chemical tagging have shown promising results \citep[e.g.,][]{de_silva07, mitschang14, hogg16, kos18}. Yet the usefulness of future chemical tagging efforts depends on each stellar population having a unique chemical abundance pattern, and it is unclear exactly how similar the chemistry of stars of a common origin actually are. Various studies have shown that the abundances of stars in open clusters are typically consistent to within $\sim$0.03 dex \citep[e.g.,][]{bovy16, ness18}. However, these studies and others typically focus on nearby, young, metal-rich populations. As we demonstrated in Paper I, wide binaries offer an opportunity to test chemical tagging across a wider range of ages and metallicities. 

In a recent study, \citet{simpson18} compared the abundances measured by GALAH \citep{de_silva15} for co-moving stellar pairs identified in TGAS by \citet{oh17}, finding significant abundance differences in five of eight stellar pairs. This is in stark contrast to the results from Paper I. However, in Paper I the abundance measurements were principally limited to overall metallicity (we additionally tested Mg, Al, Si, and Ti abundances, but these are rather imprecise, which prevented us from making any strong conclusions), whereas \citet{simpson18} compared the abundances of over 15 separate elements for each of their eight pairs. Note that these stellar pairs cannot be formally bound since they all have separations larger than the Jacobi (tidal) radius of 1.7 pc for solar mass stars in the solar neighborhood. While it is possible that they may be members of moving groups or ionized (but formerly bound) binaries \citep{jiang10}, follow-up analysis using data from \gaia\ DR2 \citep[of the wide binary sample in][]{andrews17} indicates that most stellar pairs at such wide separations have discrepant parallaxes and radial velocities (RVs) indicating they are the chance coincidence of unassociated stars \citep{andrews18b}. 

If the components of wide binaries were to be found to show large abundance differences, it would pose a problem for chemical tagging studies, as it suggests that at least some stars of a common origin have significantly variable abundance patterns. Therefore, we aim to compare the elemental abundances of wide-binary components in a consistent manner for multiple elements using a larger sample than explored previously. To do so, we require high-resolution spectra for both stars in a large number of wide binaries. In this work, we perform an independent search for the correlated motions of closely separated stars (angular separations less than 2\amin), which is indicative of binarity, using proper motions from the UCAC5 astrometric catalog \citep{zacharias17} and RVs from the Apache Point Observatory Galactic Evolution Experiment spectroscopic survey (APOGEE). Based on data taken with a fiber-fed spectrograph \citep{majewski17}, the latest APOGEE data release \citep[DR14;][]{abolfathi18, holtzman18, jonsson18} contains the abundances of more than 14 elements and RVs for $\approx$263,000 stars.

In Section \ref{sec:data}, we outline the process we use to identify our sample of wide binaries in the APOGEE/UCAC5 cross-matched catalog. We analyze the abundance consistency of wide-binary components for 14 elements in Section \ref{sec:results}. The level of chemical consistency we observe has implications for chemical tagging which we discuss in Section \ref{sec:chemical_tagging}. Finally, in Section \ref{sec:conclusions} we provide some conclusions and ideas for future directions.

\section{Wide Binaries in APOGEE -- UCAC5 catalog}
\label{sec:data}

To undertake our test of chemical tagging, we require a sample of sufficiently separated binaries for which spectra can be obtained independently for each component. These spectra must have a high enough spectral resolution and signal-to-noise ratio that accurate abundances for multiple elements can be obtained. With a spectral resolution of $R$ $\approx$ 22,500, and a typical signal-to-noise ratio in excess of 100, the APOGEE survey delivers spectra of high enough quality for this test. ASPCAP, the spectroscopic pipeline for APOGEE, derives parameters for nearly all the stars observed by APOGEE \citep{garcia_perez16}. These parameters include the effective temperature (\Teff) and surface gravity (log $g$), as well as abundance measurements for more than a dozen separate elements (and as many as 19 for some well-measured stars).

APOGEE calibrates its measured log $g$ to detailed astroseismic models of giant stars in the Kepler field, while cluster stars are used to calibrate other stellar parameters \citep{holtzman15,garcia_perez16}. The latest APOGEE data release, DR14, additionally includes stellar abundance calibrations for dwarf stars based on open clusters \citep{holtzman18}, but note that log $g$ is not calibrated for dwarf stars. In Section \ref{sec:elements} we briefly discuss how this may affect elemental abundance determinations. 

In addition to elemental abundances, the APOGEE reduction features comparisons with spectral templates that provide RV measurements with precisions typically better than $\sim$0.1 km s$^{-1}$. However, to identify wide binaries robustly, precise proper motions are also required. We cross-match APOGEE stars with the UCAC5 proper motion catalog, which is calibrated to \gaia\ DR1, to obtain the proper motion of each star. Combined, this provides precise measurements of five of the six dimensions of phase space for stars in the APOGEE catalog: positions from the Sloan Digital Sky Survey \citep[SDSS;][]{york00}, RVs from APOGEE, and proper motions from UCAC5. Note that SDSS also provides several value-added catalogs with model-dependent distance estimates (the sixth phase space dimension)\footnote{ \url{http://www.sdss.org/dr14/data_access/value-added-catalogs/?vac_id$=$apogee-dr14-based-distance-estimations}}; however, our tests show that individual distances estimated in this way are often too inaccurate for our needs. 

Throughout the remainder of this section, we provide a brief overview of the procedure we follow to identify a robust sample of wide binaries. We refer the reader to Appendix \ref{sec:sample} for details. These include a description of how we cross-match the APOGEE spectroscopic and UCAC5 astrometric catalogs in Appendix \ref{sec:cross-match}, and of how we identify a set of candidate wide binaries within this sample in Appendix \ref{sec:identifying_wide_binaries}. Additionally, we discuss how we produce a sample of random alignments for comparison, and how we select a pure sample of wide binaries from our candidate pairs in Appendices \ref{sec:random_alignments} and \ref{sec:limiting_probability}, respectively.

Within the joint APOGEE-UCAC5 cross-matched catalog, we search for pairs of stars closely separated in phase space using an adaptation of the Bayesian method we previously used in \citet{andrews17} to identify wide binaries in TGAS. In \citet{andrews17}, we were able to use parallactic distances from TGAS; however, these are unavailable for most APOGEE stars. On the other hand, because APOGEE provides RVs, we are still matching stars using five of six phase space dimensions.

\begin{figure*}
\centerline{\includegraphics[width=1.0\textwidth]{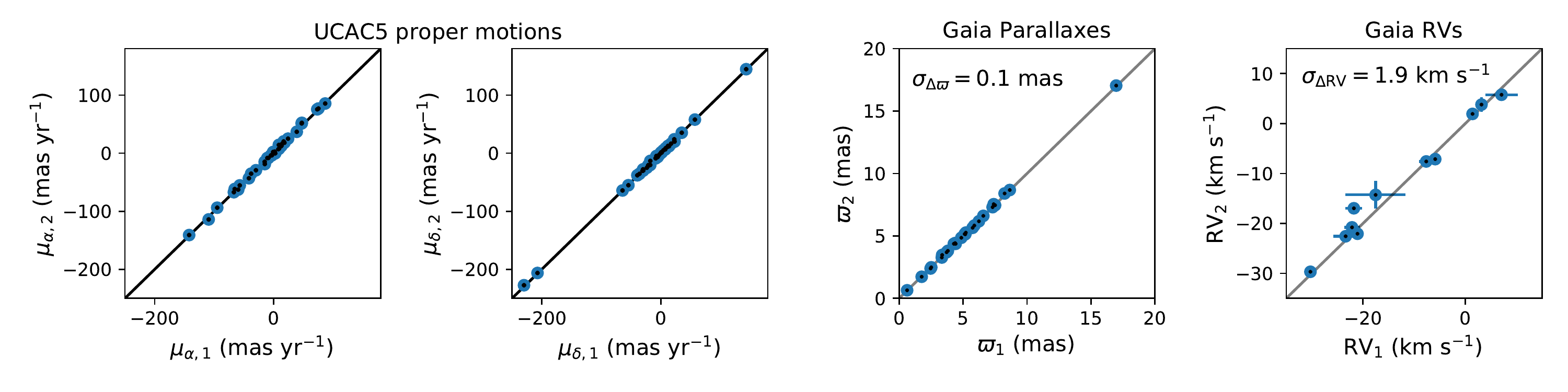}}
\caption{ We compare the UCAC5 proper motions in both right ascension ($\mu_{\alpha}$) and declination ($\mu_{\delta}$) in the left two panels. As an independent test, of the fidelity of our APOGEE sample, we cross-match the subset of our sample with posterior probability above 90\% and in \gaia\ DR2. The third panel shows that the binaries in our sample have consistent parallaxes for the 25 binaries that are measured by \gaia, while the rightmost panel shows that the binaries also have consistent RVs (except for possibly one system), for the 11 binaries with RVs measured by \gaia\ DR2. }
\label{fig:DR2}
\end{figure*}

\begin{figure}
\centerline{\includegraphics[width=1.0\columnwidth]{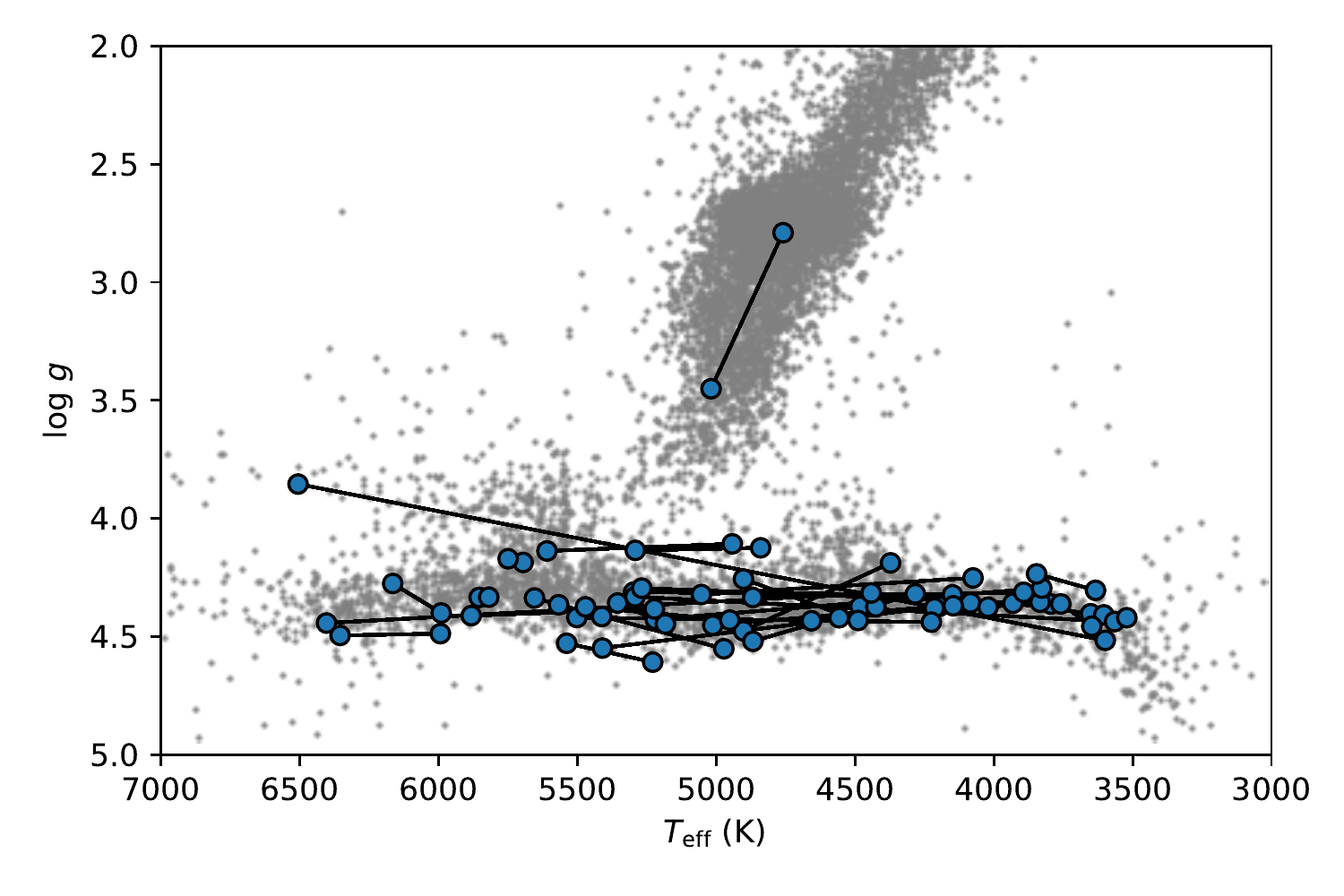}}
\caption{ The log $g$ and \Teff\ of the components in our binaries as measured by APOGEE are indicated with blue markers and linked by black lines. Gray background points show the log $g$ and \Teff\ of 10,000 random APOGEE stars. Only one of our pairs is comprised of giant stars; all others are comprised of dwarf stars. }
\label{fig:logg_Teff}
\end{figure}

\begin{figure}
\centerline{\includegraphics[width=1.0\columnwidth]{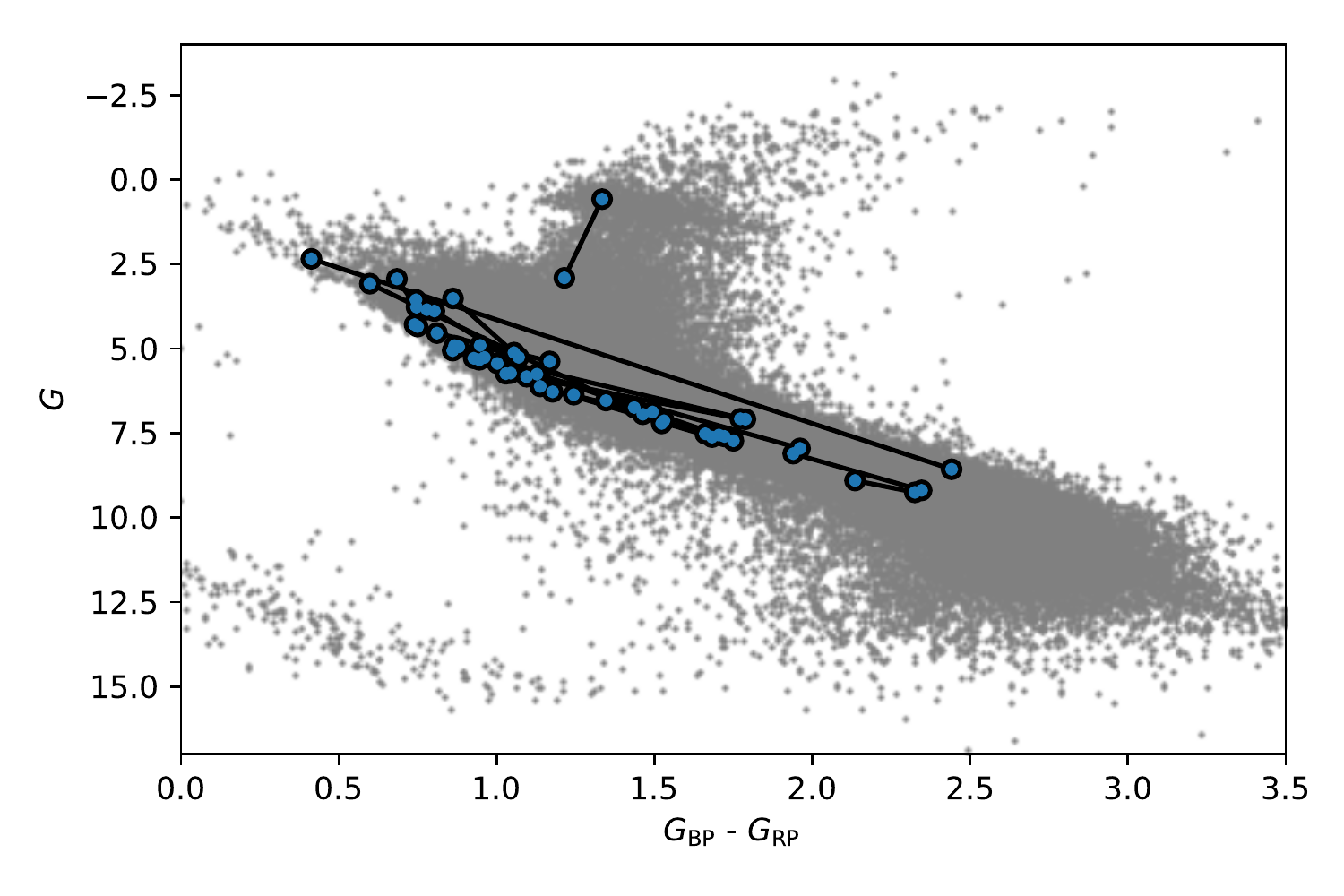}}
\caption{ The \gaia-based color-magnitude diagram for the 25 pairs in our sample with counterparts in the \gaia\ DR2 catalog. Gray points in the background show random stars from the \gaia\ DR2 catalog. }
\label{fig:CMD}
\end{figure}

In our model, each stellar pair can be assigned one of two classes, $C$: either the two stars in the pair are unassociated ($C_1$) or they form a wide binary ($C_2$). We therefore want to identify genuine binaries accurately within all $0.5 N (N-1)$ possible combinations of two stars within our APOGEE-UCAC5 cross-matched catalog. This amounts to calculating the following probability:
\begin{equation}
P(C_2 \given \vec{x}_i, \vec{x}_k) = \frac{P(\vec{x}_i \given C_2, \vec{x}_k) P(C_2 \given \vec{x}_k)}{
P(\vec{x}_i \given \vec{x}_k)}, \label{eq:P_binary_1}
\end{equation}
where we have used Bayes' Theorem to convert the posterior probability, $P(C_2 \given \vec{x}_i, \vec{x}_k)$, into calculable terms. Appendix \ref{sec:identifying_wide_binaries} provide details for how we calculate each of the terms in Equation \ref{eq:P_binary_1}.

For comparison, we also require a null sample: pairs of stars with similar kinematics that are not wide binaries. In \citet{andrews17} we were able to generate this sample by matching our parent catalog with a version of itself having shifted positions; so long as the shift is large enough, every pair found from this matching exercise will be a contaminating pair. Because APOGEE fields are relatively small (3\degree\ in diameter), shifting the positions of APOGEE stars will often move them out of the survey area. Here, we produce a sample of random alignments by selecting those pairs of stars identified by our algorithm, but with very low posterior probability: $P(C_2 \given \vec{x}_i, \vec{x}_k)<0.1$. As we show below, these stars cannot be wide binaries. At the same time, our analysis in Appendix \ref{sec:limiting_probability} suggests that we obtain a robust sample of wide binaries by selecting those stars with $P(C_2 \given \vec{x}_i, \vec{x}_k)>0.9$. None of the stars in these binaries were flagged with the ASPCAP {\tt STAR\_BAD} flag, which would indicate that the measured stellar parameters may be unreliable. 

The left two panels of Figure \ref{fig:DR2} compare the proper motion in right ascension and declination for the components of the wide binaries that we identify, those pairs with $P(C_2 \given \vec{x}_i, \vec{x}_k)>0.9$. Measurement uncertainties are smaller than the data points. That the data points lie on top of the one-to-one line is expected since the UCAC5 proper motions are included in the matching algorithm. 

As an independent test, we cross-match our resulting sample of 36 binaries with the recently released \gaia\ DR2 astrometric catalog \citep{Gaia_DR2} and find 25 candidate pairs in which both stars have parallaxes measured by \gaia. We compare the parallaxes of the stars in the 25 pairs in the third panel of Figure \ref{fig:DR2}. The right-most panel shows the subsample with RVs measured with \gaia. There is only one pair that clearly falls off the one-to-one line, which has an RV discrepancy of 4.8$\pm$1.9 km s$^{-1}$. Further epochs of RV observations are required to determine if these stars are unassociated or if one of these stars hosts a hidden companion. The clear consistency between the parallaxes and RVs of the stars in these pairs (we provide standard deviations on the differences in these quantities in Figure \ref{fig:DR2}) suggests that our sample is free from contamination by chance coincidences, in agreement with our analysis in Appendix \ref{sec:sample}.

\begin{figure}
\centerline{\includegraphics[width=1.0\columnwidth]{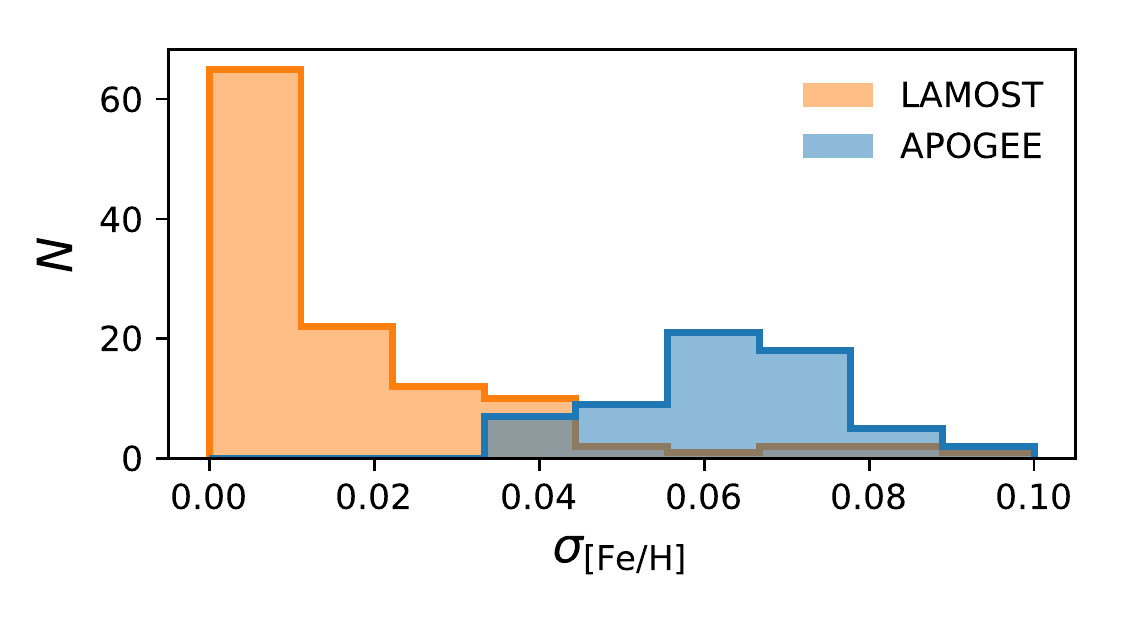}}
\caption{ APOGEE consistently measures [Fe/H] abundances with a precision $\sim$0.06 dex (blue). Although the typical LAMOST spectrum provides measurements of [Fe/H] with a precision less than 0.02 dex (orange), repeat measurements of the same LAMOST stars indicate that the real uncertainty closer to 0.06 dex \citep{luo15}. }
\label{fig:Fe_H_error_compare}
\end{figure}

\begin{figure}
\centerline{\includegraphics[width=1.0\columnwidth]{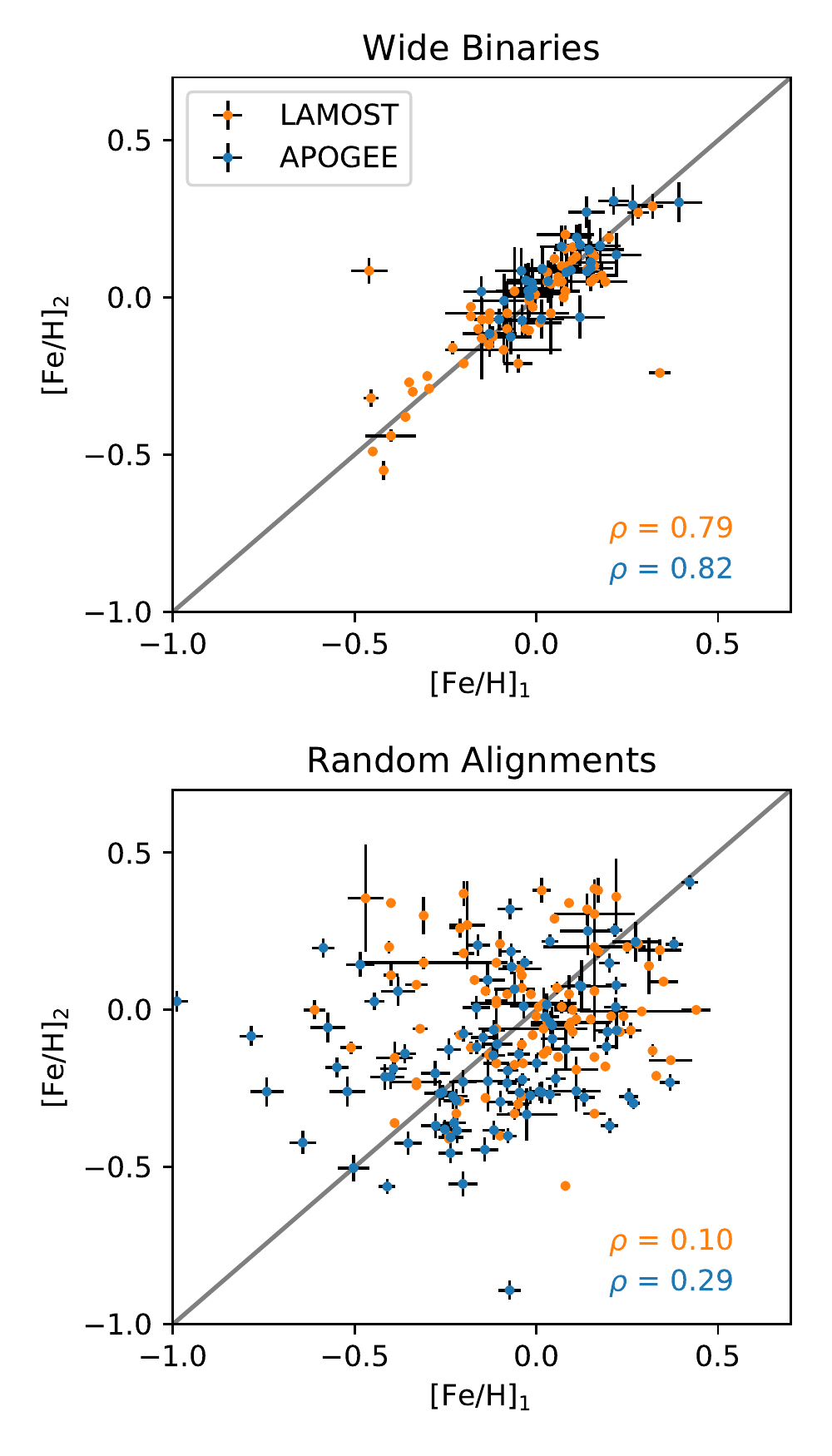}}
\caption{ We compare the [Fe/H] of each star in our wide-binary samples (top panel) as measured by APOGEE in this work and by LAMOST in Paper I. The sample presented in this work spans a narrower range of metallicities than both the LAMOST sample of wide binaries and either sample of random alignments in the bottom panel. Dramatic differences in the scatter between the top panel, which shows our wide binaries, and the bottom panel, which shows the set of random alignments, suggest that our binary samples have minimal contamination.  }
\label{fig:Fe_H_compare}
\end{figure}

In addition to abundances and RVs, the spectroscopic pipeline in APOGEE provides measurements of log $g$ (which is uncalibrated for dwarfs) and \Teff\ for the stars in our pairs, which we show in Figure \ref{fig:logg_Teff}. All but one of the wide binaries in our sample are comprised of main sequence stars. For the 25 pairs of stars that we have cross-matched using \gaia\ DR2, we show in Figure \ref{fig:CMD} the \gaia\ $G_{\rm BP}$-$G_{\rm RP}$ color against the absolute \gaia\ $G$ magnitude. Figure \ref{fig:CMD} confirms that all but two stars in our sample are on the main sequence.

APOGEE has measured stellar abundances for multiple chemical elements for 31 of the 36 binaries identified by our algorithm including the one pair of giants. These 31 binaries form the sample used throughout this work.

\begin{figure}
\centerline{\includegraphics[width=0.9\columnwidth]{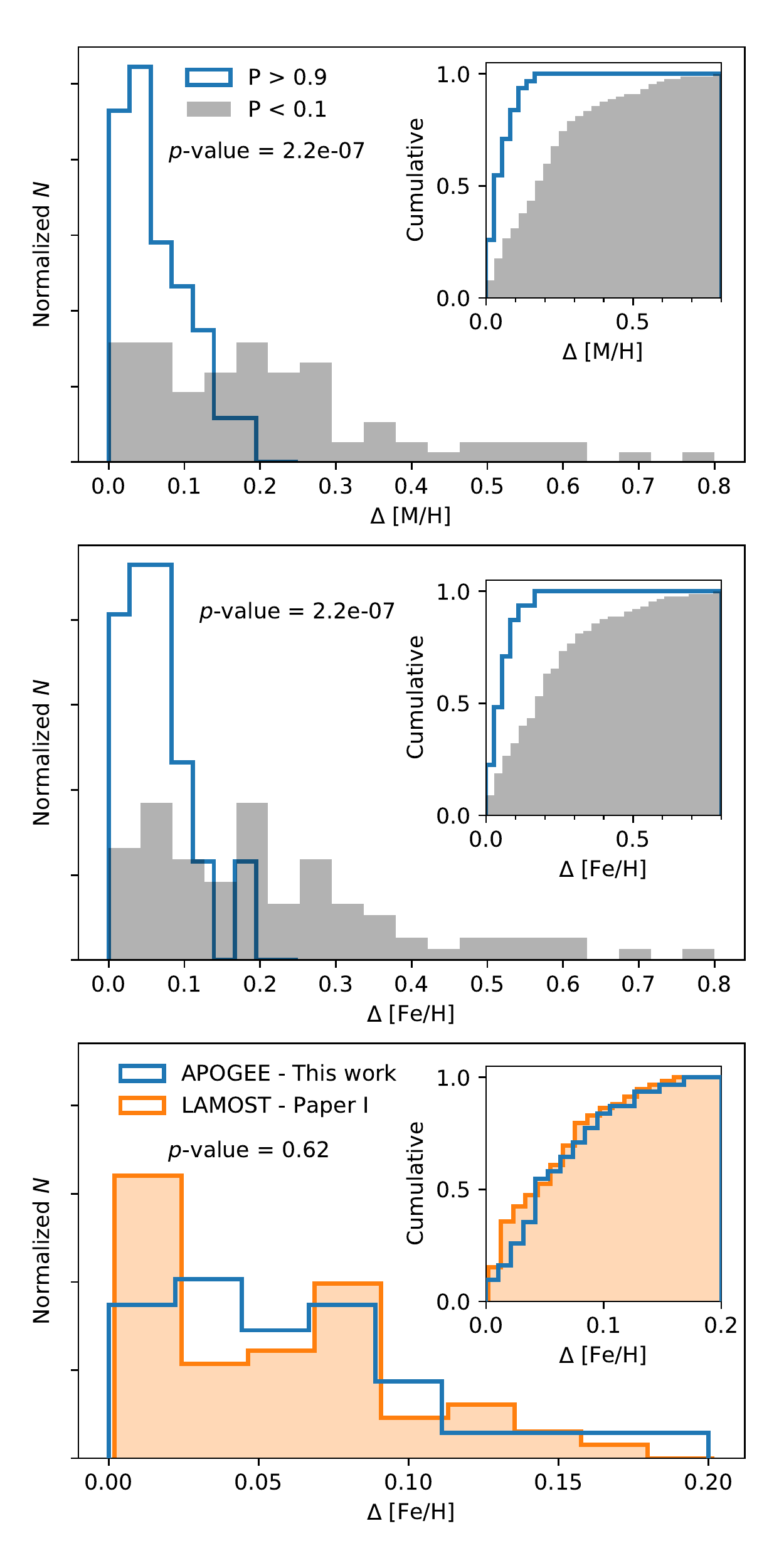}}
\caption{ For the stars in our wide-binary sample (blue), the primary metallicity indicator determined by ASPCAP \citep{garcia_perez16}, [M/H], is consistent to $\sim$0.1 dex (top panel). Fe, one of the most accurately measured elements, has a similar level of consistency within our wide-binary sample (middle panel). The chance pairings of unassociated stars (grey; generated by taking the sample of stellar pairs with $P<0.1$) show a much wider distribution in the top and middle panels, extending $\Delta$[M/H] and $\Delta$[Fe/H] to $\sim$0.6 dex. The distribution of APOGEE-measured $\Delta$[Fe/H] is similar to the distribution of [Fe/H] differences of TGAS wide binaries with LAMOST spectra (orange), as discussed in Paper I (bottom panel). Fe abundances of wide-binary components typically agree to within $\sim$0.1 dex. }
\label{fig:delta_Fe_H}
\end{figure}

\section{Results}
\label{sec:results}

\subsection{Comparison with LAMOST}
\label{sec:lamost}

\begin{figure*}
\centerline{\includegraphics[width=0.9\textwidth]{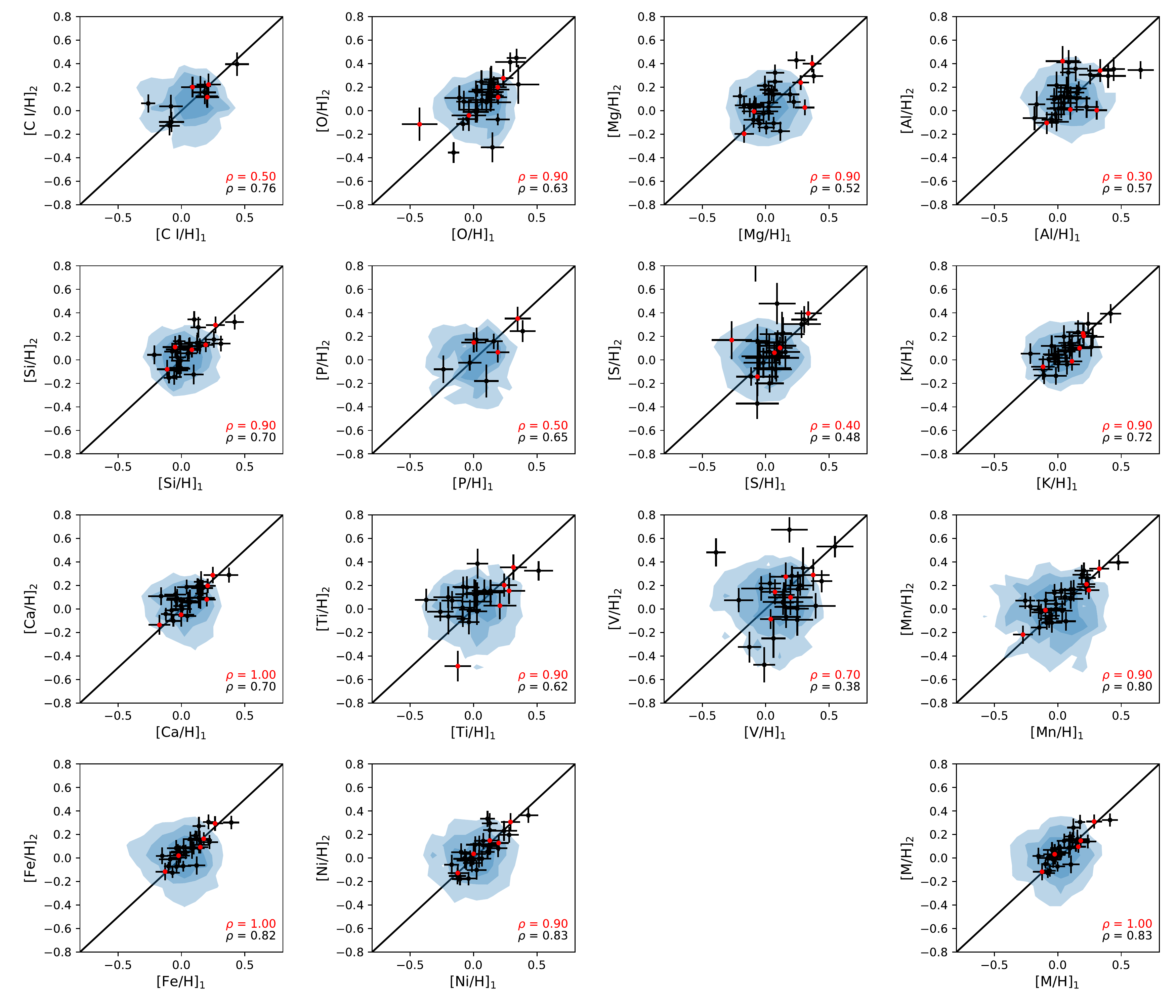}}
\caption{ The abundance differences for 14 separate elements of the stars in our sample of wide binaries (not every element is measured for every star). The bottom right panel compares the overall metallicity. Red data points indicate pairs in which ASPCAP-derived $\Delta$$\Teff<100$ K between stars in a binary pair. Spearman $\rho$ correlation coefficients for pairs with $\Delta$$\Teff<100$ K (red) as well as pairs with $\Delta$$\Teff>100$ K (black) are shown in the bottom right corner of each panel. Most panels show that selecting stars with consistent \Teff\ produces stronger correlations. Blue backgrounds in each panel compare the abundances of randomly selected APOGEE stars paired with each other.} 
\label{fig:abundance_compare}
\end{figure*}

In Paper I we compared measurements of [Fe/H] of the two stars in each wide binary in a sample identified in TGAS, but with Fe abundances measured by RAVE and LAMOST. Here, we focus on the subset of the sample from Paper I with [Fe/H] measured by LAMOST, as these stars had spectra with a very high SNR, typically above 200. In Figure \ref{fig:Fe_H_error_compare} we show that the typical LAMOST measurement uncertainties for our wide binaries are better than 0.05 dex and often as small as 0.01 dex. However, according to \citet{luo15}, the reported LAMOST uncertainties are underestimated; repeat measurements of the same stars indicate a typical [Fe/H] uncertainty of $\sim$0.06 dex \citep[see also the detailed comparison between stars measured by both LAMOST and APOGEE;][]{anguiano18}. We would like to compare the typical LAMOST measurement uncertainties with those of APOGEE.
If one uses the reported APOGEE uncertainties on [Fe/H] (\texttt{FE\_H\_ERR} column), one finds typical precisions of $\sim$0.02 dex. However, as described in \citet{holtzman18} the APOGEE DR14 Fe abundance uncertainties are known to be somewhat larger as they do not propagate uncertainties in [M/H]. Therefore, we use a more realistic formulation for [Fe/H] uncertainties: $\sigma_{\rm [Fe/H]}^2 = \sigma_{\rm [Fe/M]}^2 + \sigma_{\rm [M/H]}^2$. The blue histogram in Figure \ref{fig:Fe_H_error_compare} shows APOGEE abundance precisions in [Fe/H] for our sample to be actually of order 0.07 dex.

In Figure \ref{fig:Fe_H_compare} we show the comparison between [Fe/H] of both stars in each wide binary for both the LAMOST sample and our APOGEE binaries identified here. We provide Spearman $\rho$ correlation coefficients in the bottom right corners of both panels (a value of unity corresponds to perfect correlation, while zero is equivalent to no correlation). The top panel shows a strong correlation (with the exception of two LAMOST data points, which we suspect are contaminants; see discussion in Paper I). The bottom panel shows the corresponding sample of random alignments. The smaller scatter of the [Fe/H] in our wide-binary sample as compared to the sample of random alignments provides further evidence that our wide-binary sample is likely free of contamination.

\begin{table*}
\begin{center}
\caption{Standard deviations of the distribution of observed abundance differences. We compare the results for our sample of 31 binaries along with the subsample of our binaries showing similar \Teff. In the next two columns, we compare the abundances of the stars in our random alignments sample. We additionally provide the abundance differences for 25 of our 31 binaries as determined by {\it The Payne} \citep{ting18}. In the next two columns, we compare with the abundance differences derived from repeated observations of the same dwarf stars by APOGEE, as well as $\sqrt{2}$ median uncertainties of the 62 stars in our sample of 31 wide binaries. Finally, in the right-most column we provide the abundance consistency of open cluster stars as determined by \citet{bovy16}. \label{tab:delta_abundance} }
\begin{tabular}{lcccccccccccc}
\toprule
  & \multicolumn{2}{c}{ASPCAP} & & \multicolumn{2}{c}{Random Alignments} & & \multicolumn{2}{c}{\it The Payne} & & ASPCAP & $\sqrt{2}$ $\times$ Median & Open \\
  \cline{2-3}
  \cline{5-6}
  \cline{8-9}
 Element & All & $\Delta$$\Teff<100$ K & & All & $\Delta$$\Teff<100$ K & & All & $\Delta$$\Teff<100$ K & & Repeats & Uncertainty & Clusters \\
\bottomrule
$\Delta$$[$C\ I/H$]$ & 0.12 & 0.08 & & 0.33 & 0.32 & & $\cdots$ & $\cdots$ & & 0.09 & 0.13 & $\cdots$ \\
$\Delta$$[$O/H$]$ & 0.15 & 0.14 & & 0.32 & 0.45 & & 0.24 & 0.15 & & 0.12 & 0.13 & 0.01 \\ 
$\Delta$$[$Mg/H$]$ & 0.15 & 0.13 & & 0.27 & 0.20 & & 0.16 & 0.02 & & 0.06 & 0.10 & 0.01 \\ 
$\Delta$$[$Al/H$]$ & 0.16 & 0.22 & & 0.41 & 0.59 & & 0.13 & 0.05 & & 0.15 & 0.13 & 0.02 \\ 
$\Delta$$[$Si/H$]$ & 0.11 & 0.08 & & 0.26 & 0.18 & & 0.10 & 0.04 & & 0.06 & 0.10 & 0.01 \\ 
$\Delta$$[$P/H$]$ & 0.14 & 0.11 & & 0.43 & 0.38 & & $\cdots$ & $\cdots$ & & 0.18 & 0.13 & $\cdots$ \\ 
$\Delta$$[$S/H$]$ & 0.31 & 0.20 & & 0.26 & 0.19 & & 0.34 & 0.15 & & 0.20 & 0.17 & 0.02 \\ 
$\Delta$$[$K/H$]$ & 0.09 & 0.07 & & 0.32 & 0.37 & & 0.34 & 0.26 & & 0.19 & 0.12 & 0.03 \\ 
$\Delta$$[$Ca/H$]$ & 0.09 & 0.06 & & 0.32 & 0.41 & & 0.10 & 0.13 & & 0.08 & 0.10 & 0.02 \\ 
$\Delta$$[$Ti/H$]$ & 0.19 & 0.19 & & 0.39 & 0.49 & & 0.38 & 0.14 & & 0.21 & 0.14 & 0.03 \\ 
$\Delta$$[$V/H$]$ & 0.32 & 0.10 & & 0.46 & 0.37 & & $\cdots$ & $\cdots$ & & 0.34 & 0.15 & 0.02 \\ 
$\Delta$$[$Mn/H$]$ & 0.11 & 0.06 & & 0.40 & 0.43 & & 0.19 & 0.15 & & 0.07 & 0.11 & 0.02 \\ 
$\Delta$$[$Fe/H$]$ & 0.08 & 0.04 & & 0.33 & 0.37 & & 0.12 & 0.01 & & 0.04 & 0.09 & 0.01 \\ 
$\Delta$$[$Ni/H$]$ & 0.09 & 0.04 & & 0.32 & 0.38 & & 0.17 & 0.07 & & 0.07 & 0.10 & 0.01 \\
\bottomrule
$\Delta$$[$M/H$]$ & 0.08 & 0.04 & & 0.33 & 0.38 & & $\cdots$ & $\cdots$ & & 0.03 & 0.09 & $\cdots$ \\
\bottomrule
\multicolumn{13}{c}{Number of Pairs} \\
 \bottomrule
 & 31 & 5 & & 93 & 18 & & 25 & 3 & & $\sim$1400 & 31 & $\cdots$ \\
\bottomrule
\end{tabular}
\end{center}
\end{table*}

The top panel of Figure \ref{fig:Fe_H_compare} further demonstrates that the sample of stars in APOGEE typically have higher metallicities than the LAMOST binaries sample. This is likely because the APOGEE pointings are disk dominated and include stars at low Galactic latitudes, whereas LAMOST typically focuses on stars at higher Galactic latitudes.

\subsection{Metallicity}
\label{sec:metallicity}

We first focus on the consistency of [M/H] between component stars of our wide binaries. The top panel of Figure \ref{fig:delta_Fe_H} compares the metallicity differences between the two stars in our binaries (blue) with our random-alignments sample (grey). Our sample of wide binaries has significantly more consistent metallicities compared with our sample of random alignments. This conclusion can be seen more clearly from the inset at the top right of the top panel, which shows the cumulative distribution. Quantitatively, we provide $p$-values calculated using a two-sample Kolmogorov-Smirnov test beneath the plot legend that show the two samples are unlikely to be drawn from the same distribution at the level of nearly 10$^{-7}$. We find that [M/H] of the stars in our binaries are typically consistent to within $\sim$0.1 dex. We discuss the implications of this further in Section \ref{sec:chemical_tagging}.

\subsection{Iron}
\label{sec:iron}

We next focus on [Fe/H], as it is one of the most reliably measured elements in both LAMOST and APOGEE. The middle panel of Figure \ref{fig:delta_Fe_H} shows that our wide-binary sample (blue) shows much more consistent [Fe/H] than our random-alignments sample (grey), paralleling the top panel showing [M/H]. Again, $p$-values show the distributions are unlikely to be drawn from the same distribution at the level of $\sim$10$^{-7}$. As in the top panel, the inset in the middle panel compares the cumulative distributions. These distributions reiterate the close [Fe/H] consistency that was seen previously in Figure \ref{fig:Fe_H_compare}.

The bottom panel of Figure \ref{fig:delta_Fe_H} compares the Fe abundance differences of our wide binaries (blue) to that of the LAMOST sample (orange) from Paper I. The distribution of $\Delta$[Fe/H] is nearly identical between both samples ($p$-value of 0.65; qualitatively seen even more clearly in the cumulative distribution inset), despite the different sample selection methods, telescopes, spectral ranges, spectrographs, and analysis pipelines. As suggested by \citet{luo15}, reported LAMOST uncertainties in [Fe/H] are underestimated and ought to be $\sim$0.06 dex. Likewise, Figure \ref{fig:Fe_H_error_compare} shows that the uncertainties of the stars in our APOGEE samples (when accounting for uncertainties in [M/H]) are also $\sim$0.06 dex. It is somewhat remarkable that both LAMOST and APOGEE, experiments with two very different designs and 10$\times$ different spectral resolutions, and presumably different systematics, show a nearly identical distribution of $\Delta$[Fe/H].

\subsection{Elemental Abundances}
\label{sec:elements}

\begin{figure*}
\centerline{\includegraphics[width=1.0\textwidth]{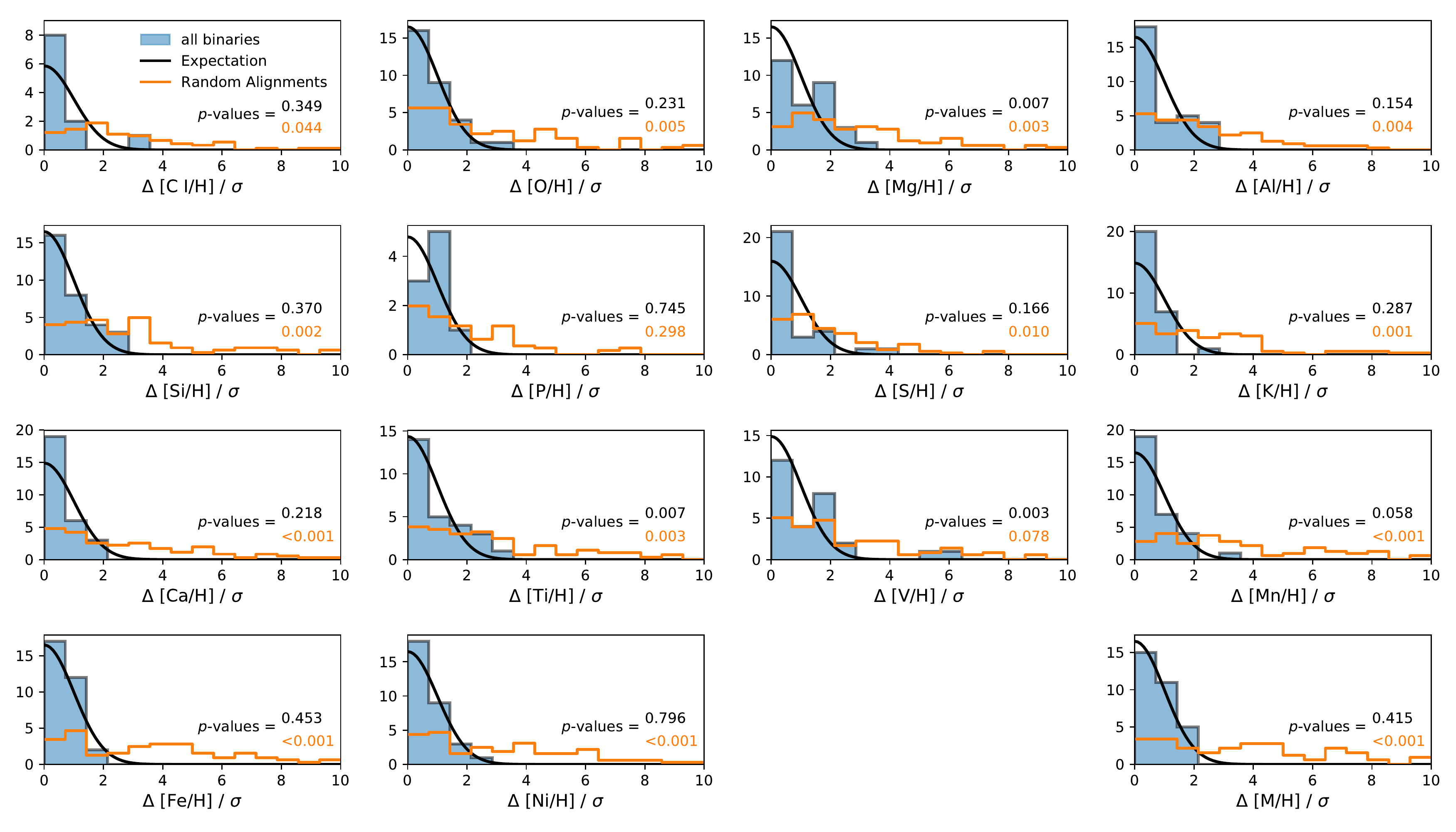}}
\caption{ We compare the abundance differences, normalized to the measurement uncertainties, of our wide-binary sample (blue) with the abundance differences of our sample of random alignments (orange). Black Gaussian curves display the expectation if the abundance differences were due to measurement uncertainties alone. In nearly every case, our distribution is less extended that our random-alignments sample but consistent with black Gaussian curves. This result is justified quantitatively by the $p$-values calculated from the two sample Anderson-Darling comparison tests for each element, comparing our wide binary sample to the expectation from uncertainties alone (black) and to our random alignments sample (orange). In addition to each of the individual elements, we show the distribution of overall metallicity differences, $\Delta$[M/H], in the bottom right panel. }
\label{fig:delta_abundance}
\end{figure*}

\begin{figure}
\centerline{\includegraphics[width=1.0\columnwidth]{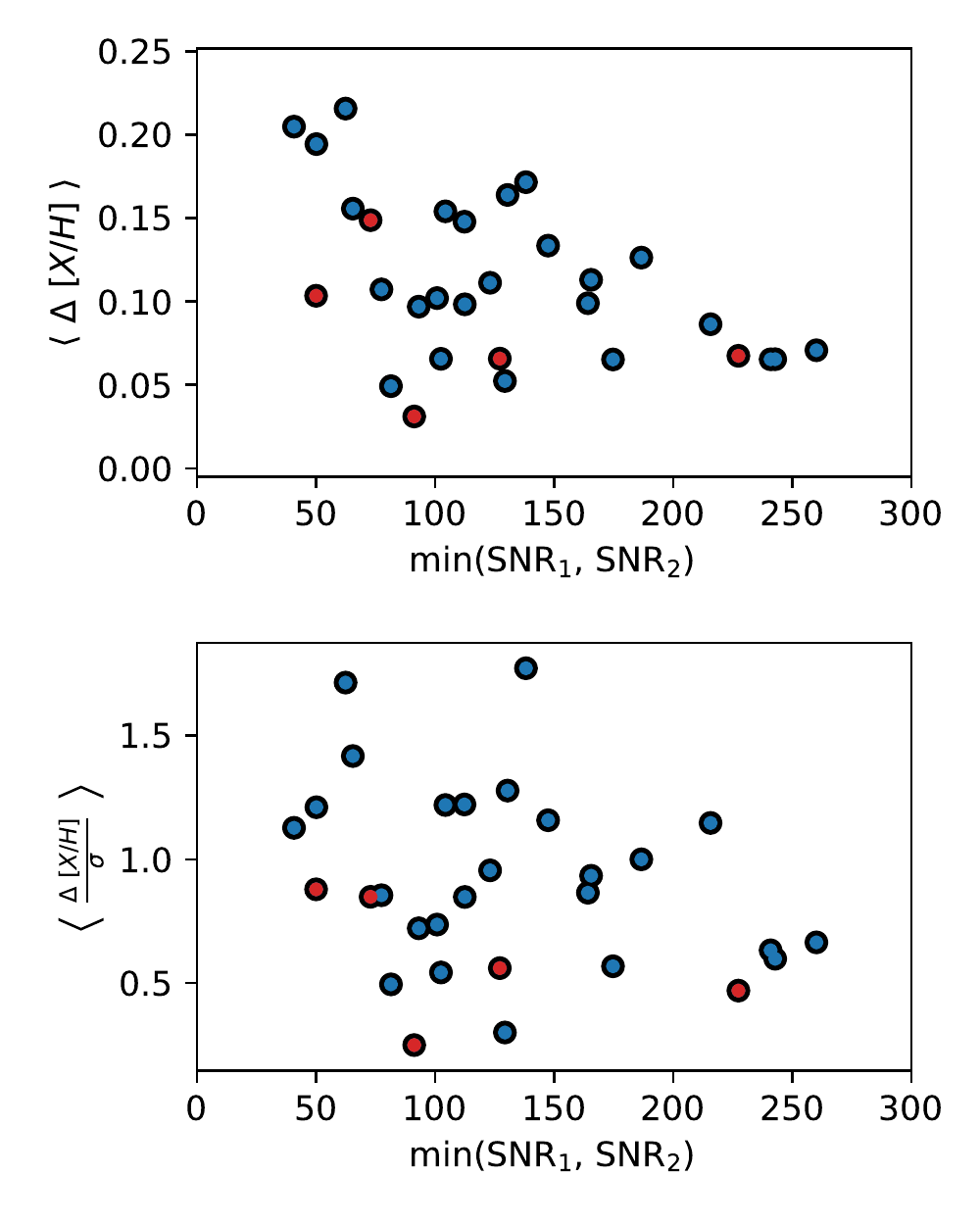}}
\caption{ The mean absolute differences (top panel) and mean differences normalized to the measurements uncertainties (bottom panel) between the components in our wide-binary sample as a function of the worst SNR of the pair. The clear trend in the top panel shows that as the SNR improves, the abundance consistency improves. Once the measurement uncertainties are taken into account, which we show in the bottom panel, this trend largely disappears (especially once two outliers are removed). Red points highlight those pairs with $\Delta$$\Teff<100$ K. }
\label{fig:delta_abundance_combined}
\end{figure}

We use the \texttt{X\_M}, \texttt{M\_H}, \texttt{X\_M\_ERR}, and \texttt{M\_H\_ERR} columns to calculate the abundances and their associated uncertainties with respect to H:
\begin{eqnarray}
{\rm [X/H]} &=& {\rm [X/M]} + {\rm [M/H]} \label{eq:abund} \\
\sigma_{\rm [X/H]}^2 &=& \sigma_{\rm [X/M]}^2 + \sigma_{\rm [M/H]}^2, \label{eq:abund_err}
\end{eqnarray}
where X is substituted for the individual elements. Throughout the remainder of this work, as elemental abundances and their uncertainties, we use the values calculated using the formulation in Equations \ref{eq:abund} and \ref{eq:abund_err}.

Figure \ref{fig:abundance_compare} expands our comparison of elements to the APOGEE-measured abundances (using the formulation above) of C I, O, Mg, Al, Si, P, S, K, Ca, Ti, V, Mn, Fe (which we repeat for comparison), and Ni. Other elements not shown here that are measured by APOGEE, such as Na and Cr, either have too few pairs with measured values, or the measurement uncertainties are too large for useful comparisons. We provide abundances for C as derived from C I lines in stars where it is measured, because in this case the adopted ASPCAP methodology applies for dwarf stars. The bottom right panel of Figure \ref{fig:abundance_compare} additionally compares the overall metallicity as measured by ASPCAP. 

Despite spanning a dynamic range of $\sim$0.5 dex, the abundances of the elements in each pair show, to varying degrees, a qualitative consistency across all of the elements. Quantitatively, Spearman $\rho$ correlation coefficients (in black) are provided in the bottom right of each panel. All elements compared have positive and relatively large correlations.

Blue backgrounds in Figure \ref{fig:abundance_compare} show the abundance differences of random APOGEE dwarfs paired together. We previously noted in Figure \ref{fig:Fe_H_compare} that the stars in our APOGEE wide binaries have [Fe/H] abundances that are larger than that of our random alignments on average. Figure \ref{fig:abundance_compare} shows that this trend extends across all elements we test: there is a consistent tail of systems with elemental abundances higher than that of random pairings of APOGEE stars. Since higher metallicity stars tend to be younger \citep{edvardsson93}, this trend could indicate that wide binaries (at least those in the Solar Neighborhood) tend to be younger \citep[a possibility discussed by][]{makarov08}. However, recent work has shown that there may not be a clearly defined age-metallicity trend among Milky Way stars \citep[e.g., ][]{casagrande11}. Further study is required to determine: (a) whether this trend remains with larger samples, (b) whether our method is more likely to identify wide binaries among high-metallicity stars, or (c) whether wide binaries are more common at higher metallicity (and possibly younger ages).

We summarize the elemental abundance differences in the second column of Table \ref{tab:delta_abundance}, where we provide the standard deviation of the abundance differences for each element for our sample of binaries. These standard deviations are calculated from the subset of our 31 wide binaries in which both stars have ASPCAP abundances for that element (typically at least 27 pairs).

Red points in Figure \ref{fig:abundance_compare} show those pairs having a \Teff\ difference between the two components less than 100 K. Quantitatively, the red correlation coefficients in the bottom right of each panel show that the abundance consistency typically improves when selecting stars with similar \Teff. The abundance differences of these pairs are quantified in the third column of Table \ref{tab:delta_abundance}.

In Figure \ref{fig:delta_abundance}, we compare the distribution of elemental abundance differences for our sample of wide binaries (blue), normalized to the measurement uncertainties, with that of our sample of random alignments (orange). The random alignment sample is scaled to show the distribution of abundance differences for our sample size. For all elements measured, the abundances of our wide pairs are more consistent than for our random-alignments sample. Table \ref{tab:delta_abundance} provides the abundance consistency of our random alignments sample, both for the entire sample (fourth column) and for the subset of our random alignments sample having \Teff\ differences between components less than 100 K (fifth column). Both columns comparing abundances of our random alignments sample show typical abundance differences 0.3-0.4 dex. With typical differences $\sim$0.1 dex, our wide binary sample has significantly more consistent elemental abundances. We provide $p$-values (in orange; calculated from an Anderson-Darling two-sample test) from comparing our wide binary sample to our random alignments sample. Most elements show $p$-values less than 0.01, which indicates that our wide binary sample shows significantly more consistent elemental abundances than our sample of random alignments.

Figure \ref{fig:delta_abundance} additionally shows black, Gaussian curves indicating the expectation of abundance differences in our sample were they produced entirely from measurement uncertainties. We provide $p$-values (in black) comparing the black, Gaussian expectation from measurement uncertainties alone, with the observed abundance differences. Most -- but not all -- elements show abundance differences that are consistent with being due to measurement uncertainties alone. The exceptions are Mg, Ti, and V. Ti and V are poorly measured elements for APOGEE, and we will not discuss these further. On the other hand, Mg is one of the best-measured elements. The consistency of Mg abundances in Figure \ref{fig:abundance_compare}, indeed, seems to show that the distribution of systems is spread beyond measurement uncertainties. When restricting to those five systems with similar \Teff, four show very close [Mg/H] consistency, but one outlier remains. It is therefore possible that wide binaries show genuine Mg abundance differences.

Alternatively, the fact that APOGEE dwarf stars have uncalibrated surface gravities could be causing some of the elemental abundance discrepancies we observe \citep{teske15}. For instance \citet{brewer15} find that different methods for measuring log $g$ result in abundances that differ by $\approx$0.06 dex. Furthermore these authors find that the differences are particularly acute for Mg abundances. Improved log $g$ measurements may therefore improve the abundance accuracy of APOGEE dwarf star samples. Further study with larger APOGEE samples is required before more definitive conclusions can be drawn.

As an independent test of the spectroscopic abundances of our sample, we additionally compute standard deviations in the abundance differences for the 25 binaries in our sample that have abundances measured by {\it The Payne} \citep{ting18}. These abundances are computed using the same APOGEE spectra used by ASPCAP, but using a physically motivated machine learning method. In Table \ref{tab:delta_abundance} we provide these values in the sixth column for all 25 pairs that cross-match and in the seventh column for the three pairs of those 25 that have $\Delta$\Teff $<$ 100 K. For most abundances, {\it The Payne} shows similar, or slightly larger, abundance differences than those measured by ASPCAP. We note that when selecting for stars with similar \Teff, the abundances as measured by {\it The Payne} become much more consistent, but one should take these results as only preliminary since these are computed for only three binaries. 

In Figure \ref{fig:delta_abundance_combined} we show these mean abundance differences (top panel) and the mean abundance differences, normalized to the measurement uncertainties (bottom panel) as a function of the minimum signal-to-noise ratio (SNR) of the APOGEE spectra. Unsurprisingly, the top panel shows that the abundance consistency improves with increasing SNR. For the few binaries in which both stars have SNR $>$200, mean elemental abundances are consistent to $\sim$0.07 dex. The bottom panel shows the mean elemental abundance differences, normalized to the measurement uncertainties. Ignoring the two largest outliers shows no strong trend exists with SNR.

\section{Implications for Chemical Tagging}
\label{sec:chemical_tagging}

\begin{figure*}
\centerline{\includegraphics[width=1.0\textwidth]{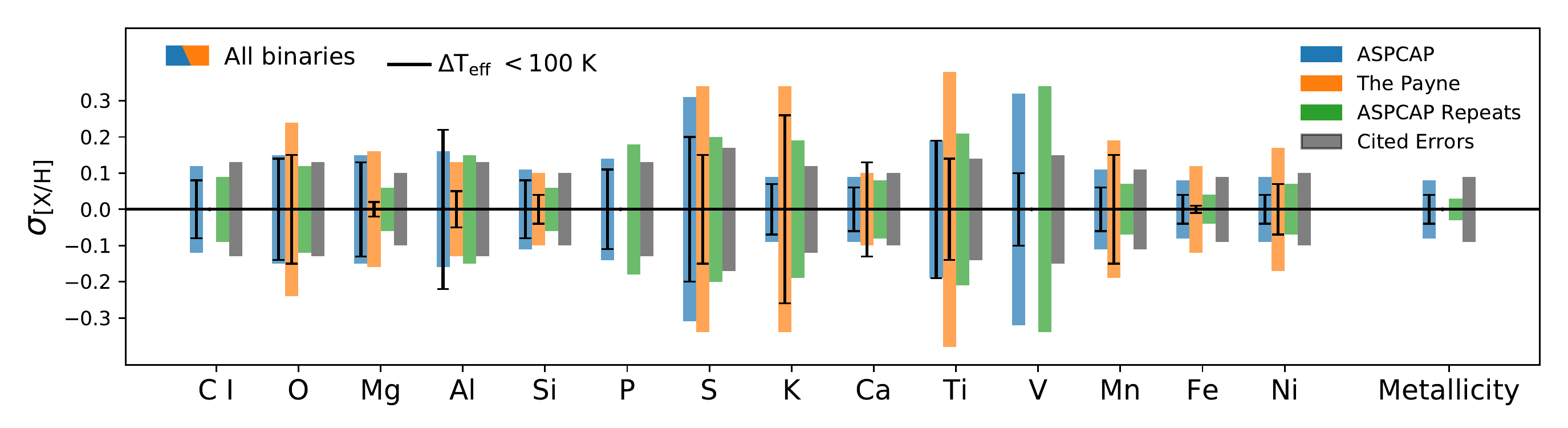}}
\caption{ We summarize the quantities in Table \ref{tab:delta_abundance}, comparing the abundance consistencies as measured by the ASPCAP pipeline (blue), {\it The Payne} (orange), and for dwarf stars with repeated measurements by ASCPAP (green). The extent of the bars corresponds to $\pm$1$\sigma$ in $\Delta$[X/H] for the stellar pairs. The distributions for all the binaries in our sample are shown as colored bars, while the subset of those binaries with $\Delta$$\Teff<100$ K are shown as black vertical lines. Selecting stars of similar \Teff\ typically improves the abundance consistency. For each element we show the median measurement uncertainties for the stars in our wide binaries as given by ASPCAP (gray). For nearly every element, the cited measurement uncertainties are very similar to the observed dispersion. At the far right, we compare the overall metallicity consistency as determined by the ASPCAP fit to atmospheric parameters. }
\label{fig:aggregate_uncertainties}
\end{figure*}

Although APOGEE is principally calibrated for giant stars, our results suggest that APOGEE dwarf stars can currently be used for chemical tagging at the level of $\sim$0.1 dex for \Teff\ ranging from 3500 K to 6500 K, an abundance consistency much better than the $\sim$0.3-0.4 dex consistency we find for our sample of random alignments. For comparison, \citet{ness18} find a level of chemical consistency for APOGEE giant stars within open clusters to be approximately 0.03 dex. There is no obvious reason why giant stars would intrinsically have significantly more consistent abundances than dwarf stars in our sample. Furthermore, focused studies of the elemental consistency of dwarf stars in clusters, such as the work by \citet{liu16} who study solar-like stars in the Hyades, find elemental variations at the level of $\sim$0.02 dex. In the tenth column of Table \ref{tab:delta_abundance}, we show the typical abundance consistency identified by \citet{bovy16} for stars in open clusters. The relatively larger abundance differences we find for our wide binaries would seem to suggest that with future improvements to ASPCAP that present a better treatment of dwarfs, better precision could be possible. 

As a test of the cited ASPCAP abundance uncertainties, we search the APOGEE catalog for stars that have been measured multiple times, but where the measurements have been processed independently throughout ASPCAP\footnote{APOGEE often takes repeat measurements of partially overlapping fields, with spectra occasionally taken for the same stars on both plates, by chance.}, as described in \citet{holtzman18}. There are $\approx$1400 such stars with log $g>4.0$ and abundance measurements for both spectra in APOGEE DR14. Since the star's intrinsic abundances will be identical between measurements, the abundance scatter in these measurements provides an assessment of random uncertainties in the ASPCAP pipeline, holding all other properties constant. The eighth column of Table \ref{tab:delta_abundance} summarizes the standard deviations of the abundance differences and shows that these differences range between 0.04 dex for Fe and 0.34 dex for V, but are typically $\sim$0.1 dex. These differences are of similar order to, but slightly more consistent than, the differences seen in our wide-binary sample. Taking [Fe/H] as an example, using repeat measurements of APOGEE stars, we find typical differences of 0.04 dex, while our binaries show differences of 0.08 dex. 

The wide binary pairs in our sample are rarely comprised of two stars with the same \Teff\ and log $g$, whereas repeat observations of the same star by APOGEE will necessarily measure the same star with identical underlying characteristics. Abundance measurements of stars with different stellar parameters are known to suffer from systematics \citep{holtzman18}. APOGEE abundances are calibrated based on dwarf stars in open clusters, but our results suggest that further improvement may be possible; the elemental consistency we observe improves for stars in wide binaries with similar \Teff. This improved elemental consistency for binaries becomes similar in magnitude to the typical abundance differences observed by APOGEE for the same stars (compare ``ASPCAP'' binaries with $\Delta$\Teff $<$100 K with ``ASPCAP Repeats" in Table \ref{tab:delta_abundance}).

In Figure \ref{fig:aggregate_uncertainties}, we graphically display the abundance consistencies of our wide binaries from Table \ref{tab:delta_abundance}. We show abundance consistencies of our wide binaries as measured by ASCAP (blue) and {\it The Payne} (orange). We additionally show the typical abundance difference observed for the same APOGEE star observed during different epochs on different plates (green) as well as the median ASPCAP uncertainty measured for our wide binaries (gray). We overplot the subset of our binaries with $\Delta$\Teff $<$ 100 K as black error bars on top of the blue and orange, solid bars.

In addition to the continuing improvements to the ASPCAP pipeline with each successive data release \citep[see discussion in][]{holtzman18}, efforts are underway to reduce the systematic effects present in spectroscopic data using both data-driven methods \citep[{\it The Cannon};][]{ness15} and physically motivated methods \citep[{\it The Payne};][]{ting18}. Currently, {\it The Cannon} is only calibrated for giant stars, while {\it The Payne} has no such limitation. Table \ref{tab:delta_abundance} and Figure \ref{fig:aggregate_uncertainties} show that for our sample {\it The Payne} offers no significant improvement over the abundances measured by ASPCAP, and in many cases the abundances are somewhat worse. There is limited evidence that the abundances as measured by {\it The Payne} are particularly consistent for stars with similar \Teff, but one should take these results as only preliminary since these are computed for only three binaries.

Until further improvements are made, chemical tagging with APOGEE dwarf stars across a range of \Teff\ seems to reach a precision of $\sim$0.1 dex. Experiments requiring more precise abundance consistencies may find that stars of different origins overlap in abundance space. This phenomenon was described by \citet{ness18} as chemical ``doppelg{\"a}ngers'': unassociated stars that have very similar chemistry. These authors find that 0.3\% of field giant stars have abundances consistent with at least one unassociated field star at the level of 0.03 dex across 20 separate elements. For dwarf stars with 14 measured elements at a precision of 0.1 dex, the chemical fingerprints that are meant to distinguish stars formed in different subpopulations may not be very discerning. We suspect that the current level of abundance precision may be enough to supplement astrometry and kinematics in identifying Milky Way subpopulations, but are likely not yet enough for identifying subpopulations based on chemistry alone.

If further precision with dwarf stars is required, our results indicate that one ought to apply chemical tagging algorithms to stars with a similar \Teff, as argued by \citet{dotter17}. When doing so, we find an elemental abundance consistency approaching 0.05 dex for the best measured elements: Fe, Si, K, Ca, Mn, Fe, and Ni. One should view these quantitative results with caution, as these are determined using only five pairs with $\Delta$ \Teff $<$100 K, but these first results are encouraging. Future tests using larger samples of APOGEE dwarf stars in open clusters and wide binaries will help to better characterize the abundance consistency of APOGEE dwarf stars of a common origin.

\section{Conclusions}
\label{sec:conclusions}

By cross-matching APOGEE stars with UCAC5 astrometry, we produce a joint catalog of Galactic stars with proper motions, RVs, and stellar abundances for at least 14 elements. We adapt the algorithm we previously used to identify wide binaries within TGAS \citep[using positions, proper motions, and parallaxes;][]{andrews17} to identify wide binaries in the cross-matched APOGEE--UCAC5 catalog (using positions, proper motions, and RVs). After using multiple methods to ensure the fidelity of our sample, including examining \gaia\ DR2 parallaxes, we obtain a sample of 31 wide binaries in the APOGEE/UCAC5 catalog.

After comparing as many as 14 separate elements, we find the abundances of our wide binaries are consistent to the level of $\sim$0.1 dex, much more consistent than the 0.3-0.4 dex consistency we see for our sample of random alignments. For all elements tested except Mg, Ti, and V, the abundance differences we observe are consistent with being due to measurement uncertainties, reaffirming our argument from Paper I (based on [Fe/H] and metallicity only) that wide binary components have a common chemical nature. Our results here contradict those of \citet{simpson18}, who find significant chemical differences between the components of five of the eight co-moving stellar pairs that they study; however, as previously mentioned, their sample contains unbound stellar pairs that are unlikely to have a common origin. Therefore, our results suggest that APOGEE dwarf stars may currently be used for chemical tagging at the level of $\sim$0.1 dex. An improved consistency of $\sim$0.05 dex is reached when using stars of similar \Teff\ and selecting only the best measured elements: Fe, Si, K, Ca, Mn, and Ni.

Although we find that wide binaries have similar abundances across multiple elements -- too similar to have anything other than a common origin -- the abundance measurements have not yet reached the level of consistency previously observed for dwarf stars in open clusters \citep[e.g.,][]{liu16}. Much effort has been spent improving the abundance measurements for giant stars in APOGEE; however, our results suggest that further improvements for the abundance determinations for dwarf stars may be possible.

Depending on the cause of the elemental abundance differences we observe, there may be serious implications for the ability of chemical tagging to identify dwarf stars of a common origin at a higher precision than the 0.05 dex level reached here. If there is an intrinsic variation in the chemistry of stars of a common origin, then different substructures in our Galaxy may not have a truly unique chemical fingerprint. Detailed studies of open clusters show that intrinsic differences are of order 0.02-0.05 dex \citep[e.g.,][]{liu16,gao18}. If, on the other hand, the abundance differences seen here are due to mixing processes \citep{dotter17,souto18} or the accretion of rocky bodies \citep{gonzalez97, schuler11a, mack14,oh18}, chemical tagging with current databases may only be effective if stars with a similar temperature are used or if one relies on only elements with low condensation temperatures. If the abundance differences are principally due to uncertainties in the abundance measurements, larger samples of co-chemical stars such as wide binaries can help with these calibrations and thereby improve elemental abundance measurements. 

The recently released \gaia\ DR2 catalog now contains astrometry for 10$^9$ stars, and wide binary samples identified within this much larger catalog \citep[e.g.,][]{el-badry18a} can be cross-matched with current and future data releases from large-scale spectroscopic surveys \citep[e.g.,][]{el-badry18b}. These samples will allow for an extension of the work we present here, using larger samples, with more elements identified across a broader range of wavelengths.

\section*{Acknowledgements}

We thank the anonymous referee for valuable comments.

J.J.A. acknowledges funding from the European Research Council under the European Union's Seventh Framework Programme (FP/2007-2013)/ERC Grant Agreement n. 617001.

B.A., H.L., C.H., and S.R.M.\ acknowledge support from NSF grant AST1616636.

J.C. acknowledges support from CONICYT project Basal AFB-170002 and by the Chilean Ministry for the Economy, Development, and Tourism’s Programa Iniciativa Científica Milenio grant IC 120009, awarded to the Millennium Institute of Astrophysics.

Funding for the Sloan Digital Sky Survey IV has been provided by the Alfred P. Sloan Foundation, the U.S. Department of Energy Office of Science, and the Participating Institutions. SDSS-IV acknowledges support and resources from the Center for High-Performance Computing at the University of Utah. The SDSS web site is www.sdss.org.

SDSS-IV is managed by the Astrophysical Research Consortium for the Participating Institutions of the SDSS Collaboration including the Brazilian Participation Group, the Carnegie Institution for Science, Carnegie Mellon University, the Chilean Participation Group, the French Participation Group, Harvard-Smithsonian Center for Astrophysics, Instituto de Astrof\'isica de Canarias, The Johns Hopkins University, Kavli Institute for the Physics and Mathematics of the Universe (IPMU) / University of Tokyo, Lawrence Berkeley National Laboratory, Leibniz Institut f\"ur Astrophysik Potsdam (AIP),  Max-Planck-Institut f\"ur Astronomie (MPIA Heidelberg), Max-Planck-Institut f\"ur Astrophysik (MPA Garching), Max-Planck-Institut f\"ur Extraterrestrische Physik (MPE), National Astronomical Observatories of China, New Mexico State University, New York University, University of Notre Dame, Observat\'ario Nacional / MCTI, The Ohio State University, Pennsylvania State University, Shanghai Astronomical Observatory, United Kingdom Participation Group, Universidad Nacional Aut\'onoma de M\'exico, University of Arizona, University of Colorado Boulder, University of Oxford, University of Portsmouth, University of Utah, University of Virginia, University of Washington, University of Wisconsin, Vanderbilt University, and Yale University.

\bibliographystyle{aasjournal}
\bibliography{references}

\appendix

\section{Identifying our Sample}
\label{sec:sample}

\subsection{Cross-Matching APOGEE with UCAC5}
\label{sec:cross-match}

The UCAC5 astrometric catalog was produced by performing new reductions of the US Naval Observatory CCD Astrograph Catalog, using stars from TGAS as the reference catalog \citep{zacharias17}. By cross-matching with \gaia\ DR1, the UCAC5 catalog contains improved proper motions for 107 million stars with typical accuracies of 1-2 mas yr$^{-1}$ ($R$=11-15 mag) and degrading to $\sim$5 mas yr$^{-1}$ at $R=$16.

\vspace{-0.5 cm}
\begin{figure}[ht]
\begin{center}
\includegraphics[width=1.0\columnwidth,angle=0]{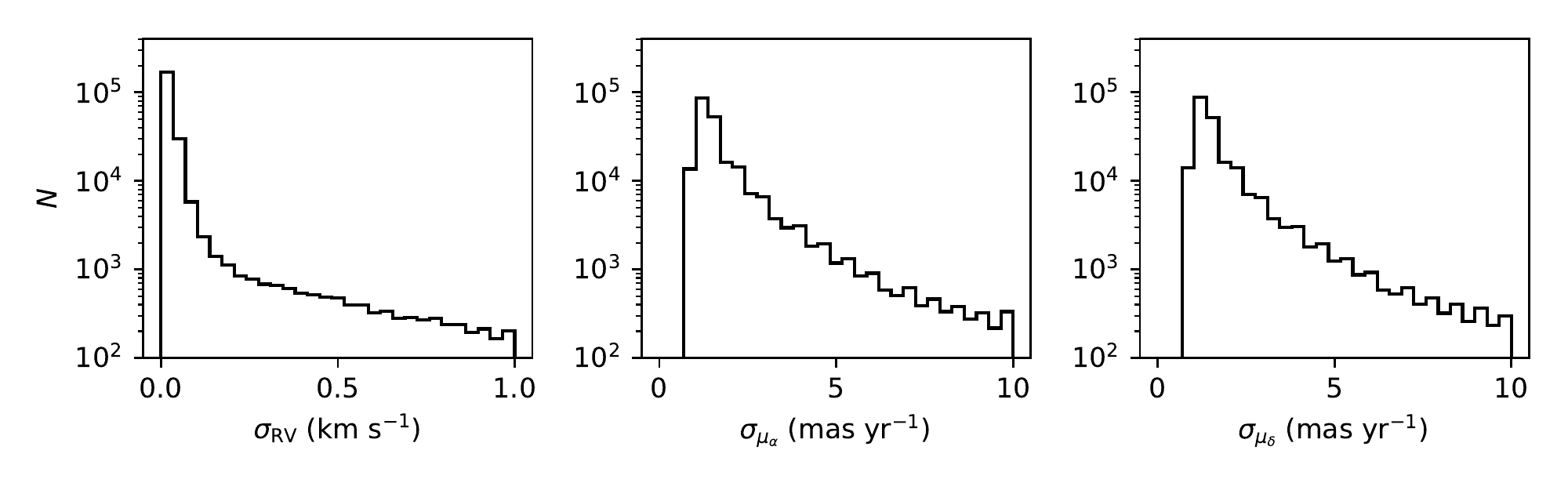}
\end{center}
\vspace{-0.5 cm}
\caption{Precisions of the joint APOGEE/UCAC5 catalog in RV(left), proper motion in right ascension (center), and proper motion in declination (right).} \label{fig:APOGEE_UCAC5_RV_ppm}
\end{figure}

Using a comparison of positions in equatorial coordinates between the surveys, we selected stars where ($\cos^2 \delta \Delta\alpha^{2}$ + $\Delta\delta^{2})^{1/2}$ $<$ 3$\asec$, and find a total of 223,754 stars in common between APOGEE DR14 and UCAC5. In Figure~\ref{fig:APOGEE_UCAC5_RV_ppm} we have the uncertainty distributions for the APOGEE/UCAC5 catalog in RV and proper motion. We find that the typical RV precisions are better than 0.1 km s$^{-1}$, while typical proper motion precisions are 1-2 mas yr$^{-1}$.

\begin{figure}[ht]
\begin{center}
\includegraphics[width=1.0\columnwidth,angle=0]{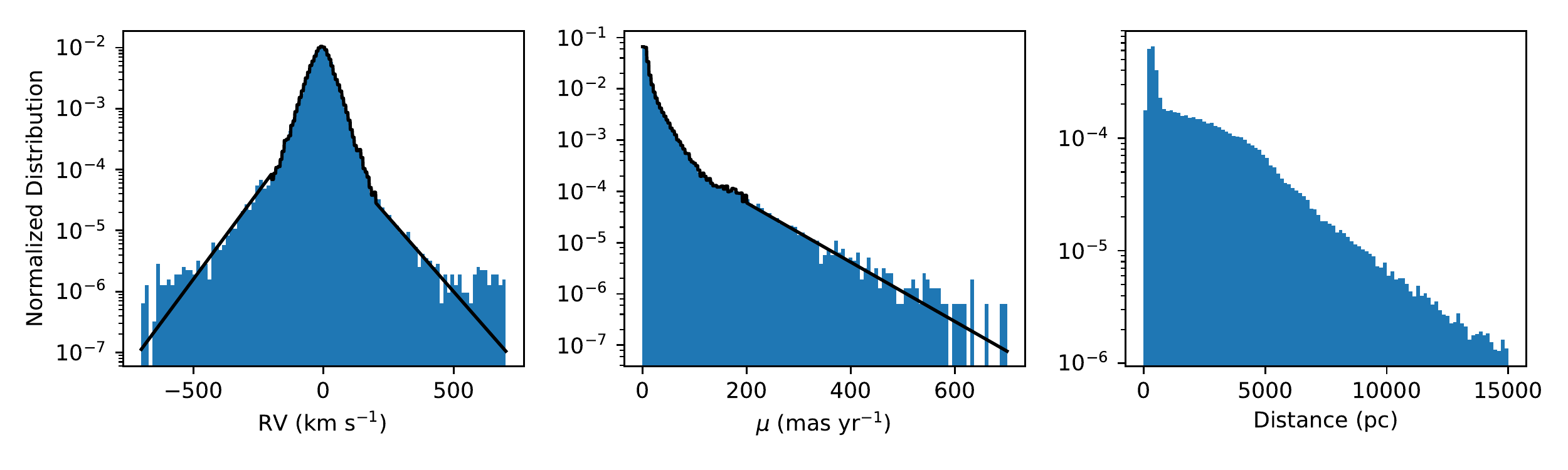}
\end{center}
\vspace{-0.5 cm}
\caption{Distribution of the joint APOGEE/UCAC5 catalog for RVs (left), proper motions (center), and APOGEE spectro-photometric distances (right). Black lines on the RV and proper motion panels show the fits to these distributions used in the wide binary identification algorithm. }
\label{fig:APOGEE_UCAC5_characteristics}
\end{figure}

We also show the distribution of the RVs, proper motions, and spectro-photometric distances \citep[provided by a value-added APOGEE catalog, calculated using the Bayesian algorithm described in][]{santiago16} for the APOGEE/UCAC5 catalog in Figure~\ref{fig:APOGEE_UCAC5_characteristics}. The black lines in the RV and proper motions panel show the fits used in the identification algorithm. See the following section for details.

\subsection{Identifying Wide Binaries}
\label{sec:identifying_wide_binaries}

Our method for identifying wide binaries within our joint APOGEE/UCAC5 catalog is based on the Bayesian method from \citet{andrews17}. The method, as we apply it here, uses the positions ($\alpha$, $\delta$), proper motions ($\mu_{\alpha}$, $\mu_{\delta}$), and RVs of stars within the catalog to calculate the likelihood that any pair of stars is either a genuine binary ($C_2$) or a chance coincidence of unassociated stars ($C_1$). We begin by defining the angular separation ($\theta$) and proper motion difference ($\Delta \mu$) between two stars using the small angle approximation:
\begin{eqnarray}
\theta &\approx& \sqrt{(\alpha_1 - \alpha_2)^2 \cos \delta_1 \cos \delta_2
                         + (\delta_1 - \delta_2)^2} \label{eq:theta} \\
\Delta \mu &\approx& \sqrt{(\mu^*_{\alpha, 1} - \mu^*_{\alpha, 2})^2 
                        + (\mu_{\delta, 1} - \mu_{\delta, 2})^2}, \label{eq:proper_motion}
\end{eqnarray}
where $\mu^*_{\alpha, i} = \mu_{\alpha, i}\cos \delta_i$. Numeric subscripts denote values associated with each of the two stellar components. 

Each stellar pair in APOGEE is defined by ten observables (and their associated measurement uncertainties) corresponding to five phase space dimensions (positions, proper motions and radial velocities) for each of two stars. We reduce the dimensionality of each stellar pair to seven, which we contain within two sets, $x_i$ and $x_k$:
\begin{eqnarray}
x_i &\equiv& \{ \theta, \Delta \mu', {\rm RV}'_1, {\rm RV}'_2 \} \\
x_k &\equiv& \{ \alpha, \delta, \mu \},
\end{eqnarray}
where primed quantities denote observed quantities with measurement uncertainties that must be accounted for.

We calculate the posterior distribution as defined by Equation \ref{eq:P_binary_1} in Section \ref{sec:data}. The denominator is separated in the same way as in \citet[][ cf. Equation 8]{andrews17}. This requires determining quantities that effectively serve as Bayesian priors: $P(C_1 \given \vec{x}_k)$ and $P(C_2 \given \vec{x}_k)$. To calculate these terms, we take the identical method as in \citet[][cf. Equations 9 to 14]{andrews17}. We refer the reader to that work for details. To wit, the prior for any stellar pair to be a random alignment scales with the square of the local density in phase space whereas the prior on any pair to be a wide binary scales linearly with the local stellar density in phase space. Normalization constants for these scalings are determined from the fact that there are $\approx 0.5 N^2$ total random alignments, whereas there are $f_b N$ total binaries in the APOGEE catalog, where $f_b$ is the binary fraction, which we take to be 50\%.

\subsubsection{Binary Likelihood}

We next wish to calculate the first term in the numerator, $P(\vec{x}_i \given C_2, \vec{x}_k)$. We first convolve over the actual proper motion and RV differences to account for measurement uncertainties:
\begin{eqnarray}
P(\vec{x}_i \given C_2, \vec{x}_k) &=& \int \dd \Delta \mu\ \dd \Delta {\rm RV}\ P(\vec{x}_i, \Delta \mu, \Delta {\rm RV} \given C_2, \vec{x}_k) \nonumber \\
&=& \int \dd \Delta \mu\ \dd \Delta {\rm RV}\ 
	P(\Delta \mu, \theta \given C_2)\ 
    P(\Delta {\rm RV} \given \theta, C_2)\ 
    P(\Delta \mu' \given \Delta \mu)\ 
    P(\Delta {\rm RV}' \given \Delta {\rm RV}), 
\end{eqnarray}
where we have substituted $x_i$ and $x_k$ for their components and split terms based on their independence: measured quantities are only dependent upon their underlying values (and measurement uncertainties). We have also made the approximation that there is no dependence on $\Delta$RV from $\Delta$$\mu$. The last two terms in the integrand are straightforwardly evaluated as Gaussian distributions centered on each stellar pair's measurement of $\Delta \mu$ and $\Delta {\rm RV}$, with a variance equal to the square of the measurement uncertainties. Note that $\Delta {\rm RV} = | {\rm RV}_1 - {\rm RV}_2|$. The analytic nature of these two terms allows this integral to be easily evaluated using Monte Carlo random sampling:
\begin{equation}
P(\vec{x}_i \given C_2, \vec{x}_k)\
\approx \frac{1}{N}\sum_j P(\Delta \mu_j, \theta \given C_2)\ P(\Delta {\rm RV}_j \given \theta, C_2), \label{eq:sum}
\end{equation}
where $\Delta \mu_j$ and $\Delta$RV$_j$ are $N$ random variates, drawn from Gaussian distributions defined by measurements of $\Delta \mu$ and $\Delta {\rm RV}$. We use 10$^4$ random draws to accurately evaluate the summation in Equation \ref{eq:sum}. 

Studies identifying wide binaries traditionally assume that wide-binary components must have the same proper motions and RVs, but as astrometry and RV measurements have improved in recent years, the orbital velocity of even the widest binaries can be detected \cite[cf. Figure 8 in][]{andrews17}. Because this orbital motion must be accounted for, evaluating the terms, $P(\Delta \mu_j, \theta \given C_2)$ and $P(\Delta {\rm RV}_j \given \theta, C_2)$, in Equation \ref{eq:sum} is not straightforward. Newtonian gravity provides the orbital velocity of a binary as a function of stellar masses, separation, eccentricity, and mean anomaly. For a particular orientation on the sky, the projected separation, tangential velocity, and radial velocity can be determined. We generate 10$^4$ random stellar binaries with primary masses drawn from a Salpeter IMF, a flat mass ratio distribution, a log flat distribution for the orbital separation, a thermal eccentricity distribution, and random orientations on the sky. 

To translate from physical units (projected separation, tangential velocity) to angular units (angular separation and proper motion) we convolve with the spectro-photometric distance \citep{santiago16} distribution provided by the value-added APOGEE catalog, which is shown in the bottom panel of Figure \ref{fig:APOGEE_UCAC5_characteristics}. Most APOGEE stars in our sample are located within the nearest 500 pc, but a significant tail extends to distances of several kpc. Our tests indicate that individual stellar distances are not accurate enough to include in our matching algorithm; however we assume that the bulk distribution is representative of the catalog. In practice, our results are not sensitive to the exact distance distribution with which we convolve.

To evaluate $P(\Delta \mu_j, \theta \given C_2)$ for any arbitrary $\mu$ and $\theta$, we construct a kernel density estimate (KDE) representation of the distribution of random binaries. We likewise represent $P(\Delta {\rm RV}, \theta \given C_2)$ with a KDE. Returning to Equation \ref{eq:sum}, we calculate the summand using terms for which we now have representations for:
\begin{equation}
P(\vec{x}_i \given C_2, \vec{x}_k)\
\approx \frac{1}{N}\sum_j P(\Delta \mu_j, \theta \given C_2)\ \frac{P(\Delta {\rm RV}_j, \theta \given C_2)}{P(\theta \given C_2)}, \label{eq:sum_2}
\end{equation}
where $P(\theta \given C_2)$ is calculated from the distribution of angular separations previously generated. 

Our search is limited to stellar pairs with angular separations less than 2\amin. For each of these pairs, we calculate the sum in Equation \ref{eq:sum_2} using 10$^4$ random samples of $\Delta \mu_j$ and $\Delta$ RV$_j$. The resulting quantity provides the likelihood that any particular stellar pair can be a wide binary.

\subsubsection{Random Alignments Likelihood}

In our Bayesian algorithm, we also require a corresponding likelihood to be calculated that represents the possibility for each stellar pair to be formed from two unassociated stars. As was performed when determining our binary likelihood, we account for measurement uncertainties in $\Delta \mu$ and the two radial velocities by convolving over their underlying values:
\begin{eqnarray}
P(\vec{x}_i \given C_1, \vec{x}_k) &=& \int \dd \Delta \mu\ \dd {\rm RV}_1\ \dd {\rm RV}_2\ P(\vec{x}_i, \Delta \mu, {\rm RV}_1, {\rm RV}_2  \given C_1, \vec{x}_k) \\
    &=& \int \dd \Delta \mu\ \dd {\rm RV}_1\ \dd {\rm RV}_2\ P(\Delta \mu \given C_1, \mu)\
    P(\theta \given C_1, \alpha, \delta) \nonumber \\
    & & \qquad \times\ P({\rm RV}'_1 \given {\rm RV}_1)\
    P({\rm RV}'_2 \given {\rm RV}_2)\
    P(\Delta \mu' \given \Delta \mu)\
    P({\rm RV}_1)\ P({\rm RV}_2), \label{eq:RA_1}
\end{eqnarray}
where we have separated terms based on independence: observed (primed) quantities are dependent only on their underlying values and the corresponding measurement uncertainties, the radial velocities are assumed to be independent of each other, and since the two stars are unrelated, $\theta$, $\Delta \mu$, and the two radial velocities are all uncorrelated. 

We can evaluate this triple integral with a Monte Carlo sum:
\begin{equation}
P(\vec{x}_i \given C_1, \vec{x}_k) \approx \frac{1}{N}\sum_j 
P(\Delta \mu_j \given C_1, \mu)\ 
P(\theta \given C_1, \alpha, \delta)\ 
P({\rm RV}_{1,j})\ 
P({\rm RV}_{2,j}), \label{eq:RA_2}
\end{equation}
where $\Delta \mu_j$, RV$_{1,j}$, and RV$_{2,j}$ are all $N$ random deviates drawn from Gaussian distributions centered on their observationally derived values and with standard deviations equal to the measurement uncertainties. We now discuss how each of the terms in Equation \ref{eq:RA_2} can be calculated in turn. 

$P(\Delta \mu \given C_1, \mu)$ is the likelihood of randomly finding two stars with a proper motion difference of $\Delta \mu$. This can be approximated as $2 \pi P(\mu_{\alpha}, \mu_{\delta}) \Delta\mu = P(\mu) \Delta\mu/\mu$, where $P(\mu)$ is the distribution of proper motions of stars in the APOGEE catalog. The middle panel of Figure \ref{fig:APOGEE_UCAC5_characteristics} shows the actual distribution of proper motions, while the black line shows our approximation generated by piecing together a KDE representation of the distribution for stars with $\mu$$<$200 mas yr$^{-1}$, and fitting a straight line in log-space to those stars with larger $\mu$.

Likewise, $P(\theta \given C_1, \alpha, \delta)$ provides a corresponding probability of finding stellar pairs with the observed angular separation, which in an analogous way to the previous term can be approximated as $2 \pi \theta P(\alpha, \delta)$. Here, $P(\alpha, \delta)$ is determined using a KDE representation of the stellar density across the sky. 

Finally, $P({\rm RV}_{1,j})$ and $P({\rm RV}_{2,j})$ are both priors on the radial velocities, which we formulate from the overall radial velocity distribution of APOGEE stars, shown as the blue distribution in the left panel of Figure \ref{fig:APOGEE_UCAC5_characteristics}. We approximate this distribution using the black line, which is calculated using a KDE representation of the data for stars with RVs between -200 and 200 km s$^{-1}$, and by fitting a line to the distribution in log-space outside this range.

We now have expression for all the terms to calculate the random alignment likelihood in Equation \ref{eq:RA_2}. We use 10$^4$ random samples to evaluate this integral.

With this binary likelihood, we can now evaluate the posterior probability in Equation \ref{eq:P_binary_1}. We apply our full expression for the posterior probability to all 33,786 pairs of stars in our APOGEE/UCAC5 catalog with angular separations less than 2\amin, finding 1282 with posterior probabilities above 10$^{-3}$. From this sample, we must determine the limiting probability which separates wide binaries from unassociated pairs of stars that by chance have similar kinematics.

\subsection{Forming a Random-Alignments Sample}
\label{sec:random_alignments}

\begin{figure}
\centerline{\includegraphics[width=0.5\columnwidth]{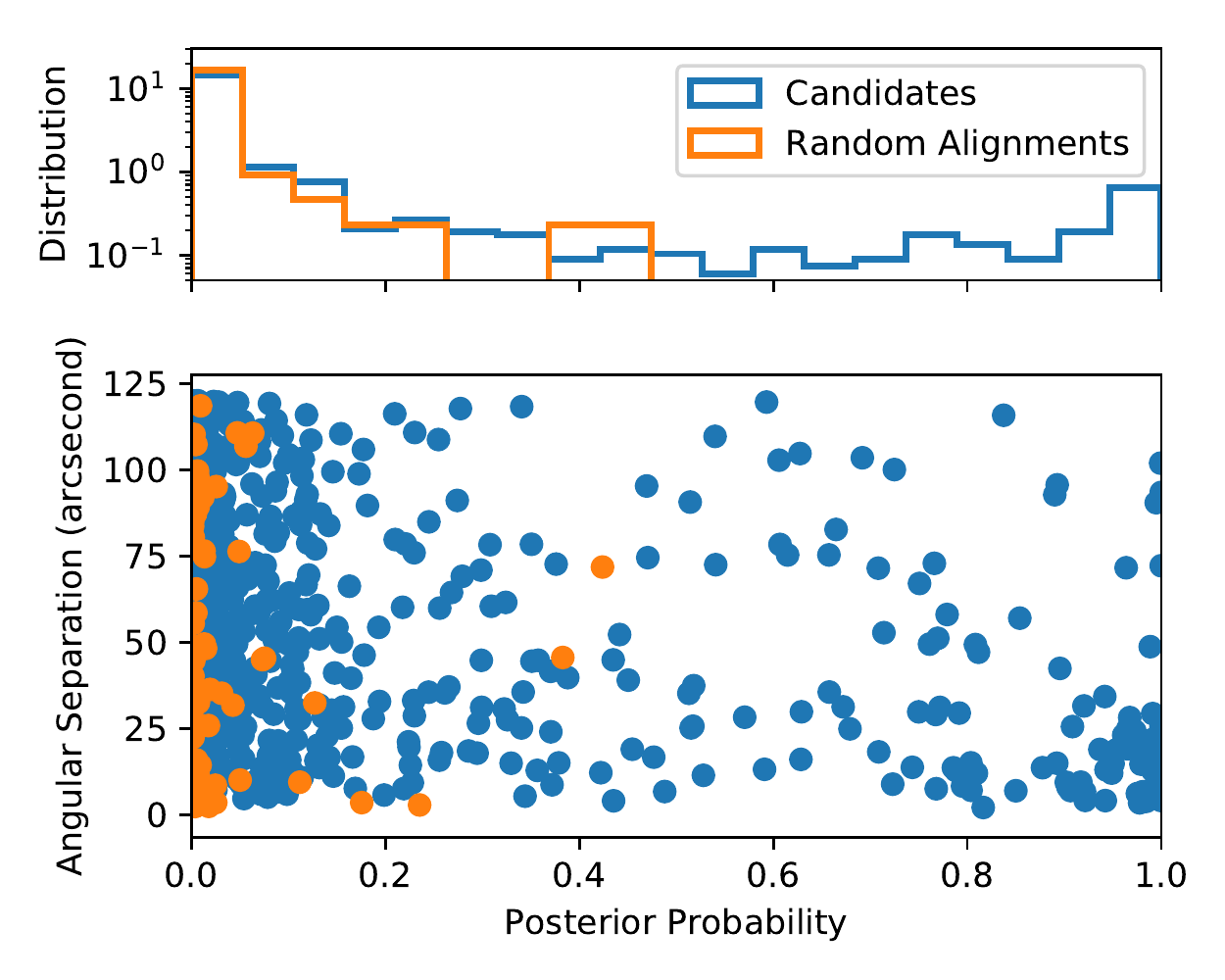}}
\caption{ The angular separation as a function of each stellar pair's posterior probability of being a wide binary for both our candidates (blue) and our random-alignments sample (yellow). The top panel shows the normalized distribution of posterior probabilities for both samples. Our method to identify random alignments underestimates the contamination fraction because the large shifts in RV and proper motion applied to the shifted catalog reduce its overlap with the original catalog; the contamination rate is actually much higher than comparison of the relative rates of blue to yellow points would suggest. Despite this, our random-alignments sample should occupy the
same parameter space as contamination in our wide binary candidates. The candidates at low posterior probability are most likely contamination due to chance coincidence, whereas the cluster of stellar pairs with posterior probabilities near unity and small angular separations are likely to be genuine. }
\label{fig:random_alignments}
\end{figure}

In \citet{andrews17}, we performed a second search in which we match the TGAS catalog with a version of itself that is shifted by $+$2\degree\ in declination and $+$3 mas yr$^{-1}$ in both proper motion dimensions. Every resulting pair identified by our algorithm is a chance coincidence. By comparing our original candidate binary sample with the catalog of chance alignments, we calibrated the limiting posterior probability that defines our sample. For our sample of APOGEE binaries, the identical test is ineffective because APOGEE is not an all-sky survey; APOGEE fields are smaller than the 2\degree\ shift we previously used. 

Instead, as a first test, we perform an analogous exercise, comparing our APOGEE-UCAC5 catalog with a version of itself shifted in declination by only $+$1\amin, but also by $+$20 km s$^{-1}$ in radial velocity and $+$5 mas yr$^{-1}$ in proper motion. We find the $+$1\amin\ shift is smaller than the APOGEE field of view, but large enough to keep most genuine binaries from artificially being identified within the shifted version of the catalog when combined with the proper motion and radial velocity shifts. Despite this, we find nine pairs from our initial catalog search that were recovered in our shifted catalog. After removing these by hand, we produce a sample of 74 random alignments with angular separations between 2\asec\ and 2\amin.

\subsection{Assessing the Limiting Posterior Probability}
\label{sec:limiting_probability}

We compare the angular separation against the posterior probability between the two samples in the bottom panel of Figure \ref{fig:random_alignments}. The top panel shows the normalized distribution of posterior probabilities. Our method of identifying random alignments underestimates the rate of chance alignments because the shifts in RV and proper motion applied to created our shifted catalog are relatively large. However, the random-alignments sample characterizes the area in phase space where contamination is higher. Both panels of Figure \ref{fig:random_alignments} show an excess of candidate pairs with posterior probabilities near unity and small angular separations, indicative of genuine wide binaries. This suggests that we would like to apply a limiting posterior probability larger than 80\%.

\begin{figure}
\centerline{\includegraphics[width=0.5\columnwidth]{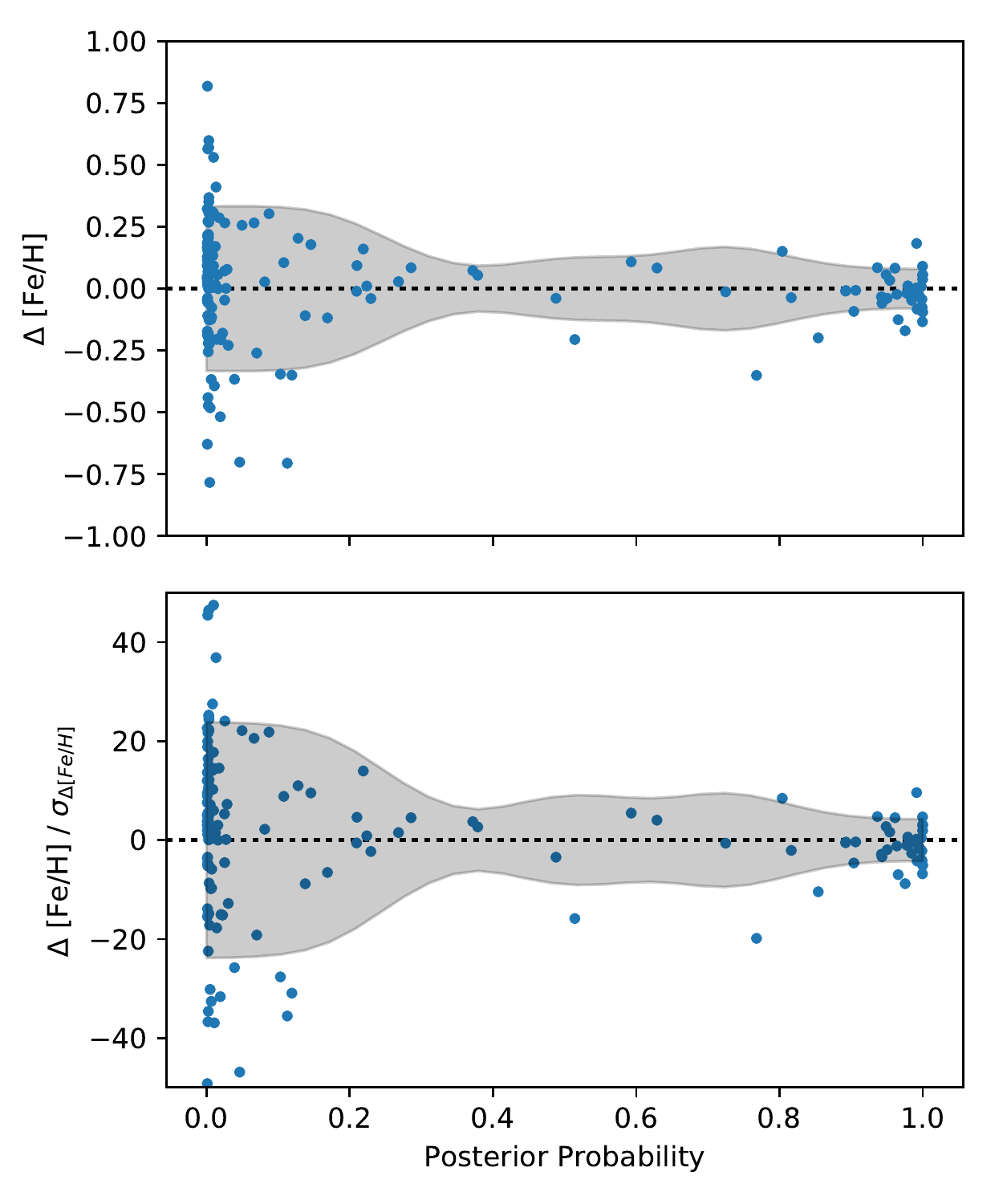}}
\caption{ We compare the posterior probability of our candidates to the metallicity difference, both absolute (top panel) and normalized to the measurement uncertainty (bottom panel), between the stellar components in each sample. That the metallicity of the pairs becomes more consistent with increasing posterior probability indicates that our statistical method is successful at identifying wide binaries. This figure serves as a consistency check, since based on our previous work in Paper I, we expect that wide binaries typically have metallicities consistent to within at least 0.2 dex. }
\label{fig:delta_metallicity}
\end{figure}

As a consistency check, we compare the Fe abundance measured by APOGEE, [Fe/H], of wide-binary components, as a function of the posterior probability assigned by our method. In paper I we determined that wide binaries have metallicities consistent to $\lesssim$0.2 dex. The top panel of Figure \ref{fig:delta_metallicity} shows that pairs at high posterior probabilities tend to have smaller Fe abundance differences, compared with those pairs at posterior probabilities closer to zero. The bottom panel shows the corresponding comparison between the posterior probability and the error-weighted metallicity difference; $y$-axis values correspond to the number of standard deviations away from zero. Grey backgrounds depict the running average in the metallicity difference, which shows a significant narrowing of the distribution as posterior probabilities increases, with a clear shift around a probabilities of 0.3.

Despite the decrease in running average at posterior probabilities of $\approx$0.3, the distribution of posterior probability and angular separation of our random-alignments sample shows that random alignments may have posterior probabilities as high as 0.4 or higher. Since we would like to obtain a relatively low contamination sample, we therefore elect to select as our sample only those stellar pairs with posterior probabilities above 0.9. Figure \ref{fig:DR2} shows that when cross-matching with \gaia\ DR2, our sample with posterior probabilities above 90\% have closely matching parallaxes. For the remainder of this work, we draw our conclusions from this sample.

\end{document}